\documentclass[aps,pra,reprint,superscriptaddress,notitlepage,showpacs,floatfix,twocolumn,longbibliography]{revtex4-2}

\usepackage{bm,bbm,amssymb,amsmath,amsfonts,amsthm,mathrsfs,MnSymbol,times}
\usepackage{natbib}
\usepackage{graphicx}
\usepackage{amsmath}
\usepackage{mathtools}
\usepackage{multirow}
\usepackage{color}
\usepackage{bbold}
\usepackage[utf8]{inputenc}
\usepackage{empheq}
\usepackage{soul}
\usepackage[ampersand]{easylist}
\usepackage[normalem]{ulem}
\usepackage{braket}
\usepackage{enumerate}
\usepackage[shortlabels]{enumitem}
\usepackage{blkarray}
\usepackage{tabularx}
\usepackage{printlen}
\usepackage{siunitx}
\usepackage{xcolor}
\usepackage[dvipsnames]{xcolor}
\usepackage{gensymb}

\usepackage{hyperref}
\hypersetup{
    colorlinks=true,       
    linkcolor=red,          
    citecolor=magenta,        
    filecolor=magenta,      
    urlcolor=cyan,           
    runcolor=cyan
}

\usepackage{outlines}
\usepackage{enumitem}
\setenumerate[1]{label=\Roman*.}
\setenumerate[2]{label=\Alph*.}
\setenumerate[3]{label=\roman*.}
\setenumerate[4]{label=\alph*.}

\begin{document}
\title{Revealing Noise in Axial Motion Through Quantum Noise Spectroscopy on a Trapped Ion Processor}
\author{Vivian Maloney}
\affiliation{Johns Hopkins University Applied Physics Laboratory, Laurel, Maryland 20723, USA}

\author{Matthew N. H. Chow}
\thanks{Present address: HRL Laboratories, LLC, Malibu, California 90265, USA}
\affiliation{Sandia National Laboratories, Albuquerque, New Mexico 87123, USA}
\affiliation{Department of Physics and Astronomy, University of New Mexico, Albuquerque, New Mexico 87131, USA}
\affiliation{Center for Quantum Information and Control, CQuIC, University of New Mexico, Albuquerque, New Mexico 87131, USA}

\author{Leigh Norris}
\thanks{Present address: Quantinuum, Broomfield, Colorado 80021, USA}
\affiliation{Johns Hopkins University Applied Physics Laboratory, Laurel, Maryland 20723, USA}

\author{Melissa C. Revelle}
\affiliation{Sandia National Laboratories, Albuquerque, New Mexico 87123, USA}

\author{Daniel S. Lobser}
\affiliation{Sandia National Laboratories, Albuquerque, New Mexico 87123, USA}

\author{Brian K. McFarland}
\affiliation{Sandia National Laboratories, Albuquerque, New Mexico 87123, USA}

\author{Edward C. Tortorici}
\affiliation{Sandia National Laboratories, Albuquerque, New Mexico 87123, USA}

\author{Christopher G. Yale}
\affiliation{Sandia National Laboratories, Albuquerque, New Mexico 87123, USA}

\author{Susan M. Clark}
\affiliation{Sandia National Laboratories, Albuquerque, New Mexico 87123, USA}

\author{Gregory Quiroz}
\affiliation{Johns Hopkins University Applied Physics Laboratory, Laurel, Maryland 20723, USA}
\affiliation{Johns Hopkins University, Baltimore, Maryland 21218, USA}

\begin{abstract}
Native noise processes in quantum processors are often difficult to isolate because multiple error mechanisms contribute to the same measured loss of coherence. Here we use dephasing-robust quantum noise spectroscopy to identify and characterize control noise induced by axial motion in an individually-addressed trapped-ion processor. When the ion motion is transverse to the addressing beam, thermal axial motion couples to the beam profile and produces effective amplitude control noise. We show that this noise is governed primarily by the local beam curvature and appears as a low-frequency contribution to the reconstructed control-noise spectrum. By varying the ion position within the beam profile, we separate curvature-dependent axial-motion noise from curvature-independent native control noise and extract motional parameters that are otherwise difficult to access on this platform. We also apply the protocol in parallel to a four-ion register, demonstrating a spectroscopic method for simultaneous characterization of position-dependent control noise across multiple qubits. The results of this study identify beam inflection points as operating regions that suppress axial-motion-induced noise at the cost of reduced Rabi rate, as found in  PRX Quantum 3, 010334 (2022).
\end{abstract}

\maketitle

\section{Introduction}

Trapped ions are promising candidates for quantum information processing, quantum sensing, and quantum clocks \cite{Tripier2026,Ransford2025,Bonus2025,Ivanov2016,Marshall2025,Ludlow2015}.  Like all quantum systems, they suffer from environmental noise, some of which can be difficult to characterize, depending on details of the experimental setup. One example is noise caused by ion motion orthogonal to the laser-beam propagation \cite{Monroe2022}.  This type of noise, if unmeasured and uncorrected, can lead to degraded performance of quantum computers, sensors, and clocks.  Typical methods for characterizing excess ion motion involve either observing ion brightness or state transfer after a Doppler-mediated interaction with a laser beam or observing ion motion using a camera as the trapping fields are changed \cite{Berkeland1998,Roos2000,Ibaraki2011}. Since ion motion can be in any or all of the three orthogonal principal axes, ensuring the ability to measure an interaction in all three directions can lead to complications in light delivery or can result in the introduction of extra laser beams---which becomes increasingly undesirable as system components are miniaturized to enable scalable quantum systems.

In this work, we extend quantum noise spectroscopy (QNS)~\cite{AlvarezPRL2011, Bylander2011, PazPRA2017, NorrisPRL2016, PazMultiaxis2019, frey2020simultaneous, amezcua2026assessing} to characterize laser-amplitude control noise while simultaneously probing axial motional dynamics in trapped-ion chains \cite{Monroe2022}, a degree of freedom that is otherwise difficult to access experimentally. Our method enables characterization of ion motion without requiring an additional laser beam with a propagation component along the motional axis. This approach also reveals a tradeoff imposed by the laser-beam profile between achievable gate speed and sensitivity to motion-induced control noise. Operating near the beam center enables maximal Rabi rates but also increases sensitivity to axial motion through the beam curvature, whereas positioning the ion near the beam inflection point suppresses motion-induced control noise at the cost of reduced drive strength \cite{Monroe2022}.

QNS provides a natural framework for investigating how axial motion affects qubit fidelity by using a quantum system as a sensor for its own noisy environment. Originally developed for dephasing noise~\cite{Bylander2011,AlvarezPRL2011}, QNS has since been extended to control fluctuations~\cite{norris2018optimally}, multi-axis noise~\cite{paz2019extending}, and correlated noise in multi-qubit systems~\cite{Uwe2020, zhou2023crqns, amezcua2026assessing}. A central challenge in applying QNS to native noise is distinguishing observed errors arising from distinct underlying mechanisms~\cite{sung2021multi}. For systems subject to concurrent dephasing and control noise, protocols have been developed to separate these contributions in both weak-~\cite{frey2017NatComm, norris2018optimally, frey2020simultaneous} and strong-noise regimes~\cite{Maloney2023}.

Using these techniques, we show that motion-induced control noise can be used to extract the mean occupation and relaxation rate of axial motional modes. We provide both theoretical analysis and experimental demonstrations on the Quantum Scientific Computing Open User Testbed (QSCOUT)~\cite{chow2023quantum,Clark2021_qscout}, where the axial direction is perpendicular to the addressing beams and therefore inaccessible to conventional laser-based motional spectroscopy techniques.

This manuscript is organized as follows. Section \ref{sec:protocol} develops a model for control fluctuations arising from axial ion motion and derives the corresponding coherence decay within the filter-function formalism. Section \ref{sect:platform} describes the experimental platform and measurement protocol. Sections \ref{sect:decay-factors} and \ref{sect:data-analysis} present measurements of the axial-motion contribution to the coherence decay, then Section \ref{sect:CoM-characterization} compares them to the theoretical model and uses the resulting fits to characterize the motional modes. Section \ref{sect:control-spectra} reconstructs the residual native control-noise spectrum. Section \ref{sect:multi-ion} extends the analysis to characterize axial-motion-induced control noise in a four-ion chain.

\section{Axial Motion Model}
\label{sec:protocol}
While hardware and environmental noise are dominant sources of errors in trapped ions \cite{soare_experimental_2014}, these systems also suffer from anomalous heating that originates from the surface of the trap, the exact cause of which is still debated \cite{brownnutt_ion-trap_2015, teller_heating_2021, an_distance_2019}. Such heating can introduce excess ion motion that leads to reductions in both single- and two-qubit gate fidelities.  One way ion motion manifests as an error is when the motion is perpendicular to the laser beam controlling the ion.  In this case, the motion results in a variation of the ion's Rabi rate (i.e., control noise) as it moves along the beam's intensity profile either during the laser pulse or between pulses~\cite{Monroe2022}.

\subsection{Dephasing-Robust Waveforms}

To isolate these control fluctuations in the presence of significant dephasing noise, we employ a recently developed family of Dephasing-Robust (DR) control waveforms \cite{Maloney2023}. These waveforms are engineered to suppress detuning and low-frequency dephasing noise while preserving sensitivity to control noise; thus, enabling DR-QNS. In particular, the protocol eliminates coherent and low-frequency contributions to the dephasing filter function (FF), effectively combining dynamical decoupling of dephasing noise with control-noise spectroscopy. As a result, dephasing contributions are dynamically suppressed to fourth order in the Magnus expansion and are therefore neglected in the derivation below. The DR waveforms are defined by
\begin{equation}
     \Omega_\text{DR}(t)
    \equiv
    \Omega_0 \, \sin(\lambda t)
    \label{eq:magical-time-domain},
\end{equation}
where the amplitude $\Omega_0$ and control modulation frequency $\lambda$ are constrained to take particular values that enforce suppression of low-frequency dephasing noise. Namely, $\Omega_0$ and $\lambda$ must satisfy
\begin{align}
&\lambda=\frac{2\pi k}{T},\label{eq::lambdaCondition}\\
&J_0\!\left(\frac{\Omega_0}{\lambda}\right)=0,\label{eq::J0Condition}
\end{align}
where $k$ is an integer, $T$ is the total duration of the waveform, and $J_0$ is the zeroth-order Bessel function of the first kind.

\subsection{Qubit-Axial Coupling}
\label{subsec:qubit-axial-coupling}
The ion's effective Rabi rate depends on both time and its displacement $x$ from the trap equilibrium position, 
\[
\Omega^{\mathrm{eff}}(x,t)=\Omega(t)\,[1+\eta(t)]\,f({x}),
\]
where the ideal control envelope $\Omega(t)$ is modulated by a stochastic gain and the position-dependent axial gain $f(x)$. The stochastic gain $\eta(t)$ is zero-mean stationary noise with power spectral density (PSD) $S_{\eta}(\omega)$, capturing time-correlated fluctuations in the laser strength.
The function $f({x})$ is a dimensionless \textit{axial gain} which describes the local beam intensity sampled by the ion as it moves laterally through the beam profile. Because the ion undergoes thermal axial motion, its displacement is treated as the stochastic process $x(t)$, and consequently the axial gain becomes the time-dependent process $f(x(t))$.

We model the axial gain function by Taylor expanding the beam profile to second order about the ion's equilibrium position, $f(x) \propto a+b x+d x^2$ \cite{Monroe2022}. Axial motion is treated as independent of $\eta(t)$.
The Rabi rate is calibrated so that $\langle\Omega^{\mathrm{eff}}(x,t)\rangle=\Omega(t)$, or equivalently so that the mean gain satisfies \(\langle f(x)\rangle=1\). This can be done experimentally without knowing $\langle x^2 \rangle$. Within our quadratic expansion, this calibration is equivalent to dividing $f(x)$ by $a + d \langle x^2 \rangle $. In the following, we reuse \(b\) and \(d\) to denote these normalized linear and quadratic coefficients.

The total Hamiltonian includes the free qubit and axial-mode Hamiltonians together with the position-dependent control interaction,
\begin{equation}
H(t)=H_q+H_a+\frac{\hbar}{2}\Omega(t)[1+\eta(t)]\hat{\sigma}_x\otimes f(\hat{x}),
\end{equation}
where
\[
H_q=\frac{\hbar\omega_q}{2}\hat{\sigma}_z,
\qquad
H_a=\sum_m\hbar\omega_m\left(\hat{a}_m^\dagger\hat{a}_m+\frac{1}{2}\right),
\]
and the displacement operator of the addressed ion is
\begin{equation}
\hat{x}=\sum_m b_m\sqrt{\frac{\hbar}{2M\omega_m}}
\left(\hat{a}_m+\hat{a}_m^\dagger\right). \label{eq:displacement}
\end{equation}
Here, \(M\) is the ion mass, \(\omega_m\) is the frequency of axial normal mode \(m\), and \(b_m\) is the participation of the addressed ion in that mode \cite{James1998-nt}.

Because dephasing noise is dynamically suppressed by the DR control,
we omit it from the effective noise model and retain only amplitude
fluctuations arising from native control noise and qubit--axial coupling.
In the resonant qubit rotating frame and the interaction picture of the
axial free Hamiltonian, the control interaction is
\[
H_{\mathrm{c}}(t)
=
\frac{\hbar}{2}\Omega(t)\hat{\sigma}_x
\otimes\left[1+\tilde f(t)\right], 
\]
where
\begin{equation}
\tilde f(t)
\equiv
[1+\eta(t)]f(\hat{x}(t))-1 \label{eq:effective-gain-fluctuation}
\end{equation}
has zero mean. We transform into the toggling frame generated by the ideal-control evolution to remove the deterministic component. Because the
ideal and fluctuating control terms are both proportional to
\(\hat{\sigma}_x\) at all times, the remaining fluctuation Hamiltonian is
\begin{equation}  
\tilde H(t)
=
\frac{\hbar}{2}\Omega(t)
\hat{\sigma}_x\otimes\tilde f(t). \label{eq:H-tilde-f}
\end{equation}
For the DR waveforms considered here,
\[
\int_0^T\Omega(t)\,dt=0,
\]
so the ideal-control propagator returns to the identity at \(t=T\).
Final observables may therefore be evaluated in the ideal-control
interaction frame without an additional deterministic rotation.

\subsection{Axial Motion}

We describe the axial motion of the coupled ion chain in terms of its
weakly damped quantum harmonic normal modes \cite{James1998-nt}. The axial modes are additionally coupled to a thermal environment where $\gamma_m$ is the damping rate or linewidth of the axial mode-environment coupling, $\beta$ is the inverse temperature, and $n_m\equiv 1 \big/ (e^{\beta \hbar \omega_m} - 1)$ is the thermal quanta, which in the high temperature limit is approximately $1\big/\beta \hbar \omega_m$.

Under the assumptions of weak-coupling and negligible back-action, the evolution of the axial modes under the influence of the external environment can be described by a Lindblad master equation $\dot{\rho}=\mathcal{L}[\rho]$ with steady-state $\rho_{\textrm{th}}$.
We can solve the Lindbladian in the interaction picture for the annihilation operators $\dot{\hat{a}}_m = \mathcal{L}^\dag [\hat{a}_m]$ and obtain the time-dependent displacement operator $\hat{x}(t)$ [i.e., Eq.~\eqref{eq:displacement}] for $t\ge0$ with $\hat{a}_m(t)= \hat{a}_m e^{- i \omega_m t - \gamma_m t / 2}$. Full details of this calculation are given in Appendix~\ref{app:Lindblad}, which yields a correlation function 
\begin{eqnarray}
C_x&(t)&\; \equiv \langle \hat{x}(t)\hat{x}(0)\rangle \label{eq:Cxt} \\
&=& \sum_m \frac{\hbar b_m^2}{2M\omega_m}
e^{-\gamma_m \lvert t \rvert /2} \Bigl[(2n_m+1)\cos(\omega_m t)-i\sin(\omega_m t)\Bigr]. \nonumber
\end{eqnarray}

\subsection{Qubit Evolution}

Armed with the dynamics of the axial modes in the absence of qubit coupling, we now consider the joint evolution of the qubit and axial modes. The system is assumed to initially reside in a separable state $\rho(0) = \rho_q(0) \otimes \rho_{\textrm{th}}$, where $\rho_q(0)$ is the initial state of the qubit. We work within the $\ket{\pm}$ basis for the qubit as they are eigenstates of the qubit portion of Eq.~\eqref{eq:H-tilde-f} at all times; thus, 
$\rho_q(0) = \sum_{\alpha,\beta \in \pm} \rho_{\alpha\beta} (0) \ket{\alpha} \bra{\beta}
$. Applying the propagator for Eq.~\eqref{eq:H-tilde-f} and tracing out the axial modes, we obtain
\begin{equation}
\begin{aligned}
    \rho_q(T)
    &= \sum_{\alpha,\beta \in \pm}
    \rho_{\alpha \beta} (0) K_{\alpha\beta}(T)\ket{\alpha} \bra{\beta}
\end{aligned}
\label{eq:rhoT}
\end{equation}
where
\begin{eqnarray}
    K_{\alpha\beta}(T)&=& \left\langle
    \mathcal{T}_- e^{i \beta \int_0^T \frac{1}{2} \Omega(t) \tilde{f}(t) dt}
    \mathcal{T}_+ e^{-i  \alpha \int_0^T \frac{1}{2} \Omega(t) \tilde{f}(t) dt}
    \right\rangle.\quad
    \label{eq:decay-factor}
\end{eqnarray}
Here \(\mathcal T_+\) and \(\mathcal T_-\) denote time-ordering and anti-time-ordering, respectively. The expectation value is taken over the laser noise and axial modes in their thermal state. In the $\ket{\pm}$ basis, we see that the dynamics of the qubit just contribute a sign to the argument of the axial expectation value; this scalar coupling lets us evaluate the expression in the axial subsystem rather than the combined system.

The decay dynamics are fully characterized by the off-diagonal factor $K_{+-}(T)$, since the diagonal factors satisfy $K_{\alpha\alpha}(T)=1$ while the two off-diagonal terms are complex conjugates. We evaluate \(K_{+-}(T)\) by expanding the time- and anti-time-ordered exponentials to second order and retaining the second cumulant. Although the thermal displacement \(\hat{x}(t)\) is Gaussian, the quadratic axial-gain and multiplicative laser-noise terms make the effective fluctuation \(\tilde f(t)\) non-Gaussian in general. The second-order cumulant truncation therefore constitutes a weak-noise approximation and yields
\begin{align}
    \log K_{+-}(T)
    = -&\frac{1}{2} \int_0^T dt_1 \int_0^T dt_2  \Omega(t_1)\, \Omega(t_2)\, \tilde{C}(t_1- t_2)
    \label{eq:logKpm}
\end{align}
with the full calculation shown in Appendix~\ref{app:Keldysh_cumulant}.

Using that $\langle \tilde{f}(t) \rangle = 0$, the overall real symmetrized cumulant including both axial thermal and laser fluctuations is
\begin{align}
    \tilde{C}(t_1 - t_2)
    &= \frac{1}{2} \left\langle
    \left\{\tilde{f}(t_1), \tilde{f}(t_2) \right\}
    \right\rangle,
    \label{eq:C_sym}
\end{align}
which is well-defined from the time difference $\tau \equiv t_1 - t_2$ due to stationarity.

The effective cumulant function entering Eq.~\eqref{eq:C_sym} is determined directly by the laser-noise correlation and the cumulants of the axial gain
\begin{equation}
\tilde C(\tau)
=
C_\eta(\tau)
+
C_f(\tau)
+
C_\eta(\tau)C_f(\tau).
\label{eq:Ctilde_decomp}
\end{equation}
The correlation of the zero-mean laser fluctuation is $C_\eta(\tau)
\equiv
\left\langle\eta(\tau)\eta(0)\right\rangle$, and as shown in Appendix ~\ref{app:effective_gain_correlation} the axial gain cumulant $C_f(\tau)$ is related to the linear and quadratic symmetrized axial displacement cumulants in Eq.~\eqref{eq:Cxt} through
\begin{align}
C_f(\tau)
&\equiv
\frac{1}{2}
\left\langle
\left\{
f(\tau) - 1, f(0) - 1
\right\}
\right\rangle
\label{eq:Cf} \\
&=
b^2 C_x^{\mathrm{sym}}(\tau)
+
d^2 C_{x^2}^{\mathrm{sym}}(\tau).
\label{eq:Cf_axial_decomp}
\end{align}
The symmetrized linear displacement correlation $C_x^{\mathrm{sym}}(\tau)$ is the real part of the
full correlation in Eq.~\eqref{eq:Cxt}, and by applying Wick's theorem the
quadratic cumulant can be expressed as
\begin{equation}
C_x^{\mathrm{sym}}(\tau) = \operatorname{Re}C_x(\tau), \quad C_{x^2}^{\mathrm{sym}}(\tau) = 2\operatorname{Re}\!\left[C_x(\tau)^2\right].
\label{eq:Cx_sym_relation}
\end{equation}
Dropping the negligible imaginary component in the high-temperature limit, the overall cumulant in Eq.~\eqref{eq:C_sym} therefore decomposes as
\begin{equation}
\begin{aligned}
\tilde C(\tau)
={}&
C_\eta(\tau)
+b^2 C_x^{\mathrm{sym}}(\tau)
+2d^2\left[C_x^{\mathrm{sym}}(\tau)\right]^2
\\
&+
b^2 C_\eta(\tau)C_x^{\mathrm{sym}}(\tau)
+
2d^2 C_\eta(\tau)
\left[C_x^{\mathrm{sym}}(\tau)\right]^2.
\end{aligned}
\label{eq:Ctilde_explicit}
\end{equation}

We define the decay exponent appearing in Eq.~\eqref{eq:logKpm}
\begin{equation}
\chi \equiv -\log K_{+-}(T), \label{eq:chi}
\end{equation}
which governs the coherence decay and hence all measured signals in this protocol. In the following, we express $\chi$ in the filter-function formalism (FFF)~\cite{ball2015walsh,green2013arbitrary} and show that, for the DR waveforms used here, it acquires a straightforward $\Omega_0^2$ scaling.

\subsection{Spectra and Filter Functions}

To evaluate Eq.~\eqref{eq:chi} for stationary control noise, we work in the frequency domain using the FFF. We define the effective control-noise spectrum, which depends on both laser intensity fluctuations and axial motion, as the Fourier transform of the symmetrized covariance in Eq.~\eqref{eq:C_sym},
\begin{equation}
S(\omega) \equiv \int_{-\infty}^{\infty} dt\,
e^{-i\omega t}\tilde C(t).
\label{eq:psd}
\end{equation}
The corresponding control FF is defined as
\begin{equation}
F(\omega,T)=
\left|\int_0^T dt\,e^{-i\omega t}\Omega(t)\right|^2,
\label{eq:filter_definition}
\end{equation}
which for the DR waveform in Eq.~\eqref{eq:magical-time-domain} becomes \cite{Maloney2023}
\begin{equation}
F(\omega,T)
=
\left(
\frac{2\Omega_0\lambda\sin(\omega T/2)}
{\omega^2-\lambda^2}
\right)^2.
\label{eq:DR_filter}
\end{equation}
With these definitions, the cumulant in Eq.~\eqref{eq:chi} becomes
\begin{equation}
\chi
=
\frac{1}{2\pi}\int_0^\infty d\omega\,
S(\omega)F(\omega,T),
\label{eq:FFF}
\end{equation}
where the integral is restricted to positive frequencies because both
\(S(\omega)\) and \(F(\omega,T)\) are even in the high-temperature limit.

\subsection{Noise Pathways}
\label{subsec:noise-pathways}
Treating the underlying laser-amplitude fluctuation and axial
displacement as independent stationary Gaussian processes, the effective spectrum can be
decomposed from Eq.~\eqref{eq:Ctilde_explicit} as
\begin{equation}
\begin{aligned}
S(\omega)
&=S_\eta(\omega)+b^2S_x(\omega)+2d^2S_{xx}(\omega) \\
&\quad +\frac{b^2}{2\pi}(S_\eta*S_x)(\omega)
+\frac{2d^2}{2\pi}(S_\eta*S_{xx})(\omega).
\end{aligned}
\label{eq:full_noise_decomp}
\end{equation}
Here \(*\) denotes convolution, \(S_\eta(\omega)\) is the (unknown) native laser-amplitude noise spectrum, \(S_x(\omega)\) is
the symmetrized displacement spectrum, and \(S_{xx}(\omega)\) is the quadratic
axial spectrum equal to the self-convolution of \(S_x(\omega)\). Due to the form of the displacement correlation function [Eq.~\eqref{eq:Cxt}], the axial mode PSDs are proportional to the Lorentzian
\begin{align}
 L(\gamma,\omega)
    &=
    \int_{-\infty}^{\infty} dt\,
    e^{-\gamma |t|/2}e^{-i\omega t}
    =
    \frac{\gamma}{\gamma^2/4+\omega^2}.
    \label{eq:lorentzian_spectra}
\end{align}
We will make use of this function below, but note that explicit expressions for \(S_x\) and \(S_{xx}\) in terms of $L(\gamma, \omega)$ are derived in
Appendix~\ref{app:axial_spectra}. 

We now identify the terms in Eq.~\eqref{eq:full_noise_decomp} that contribute appreciably to the filter overlap in Eq.~\eqref{eq:FFF}. For the DR waveforms used here, the filter is concentrated in the kHz
band, whereas the axial mode frequencies satisfy
\(\omega_m/2\pi\sim\mathrm{MHz}\). Thus, the only axial contribution with
appreciable support near the filter passband is the \textit{diagonal difference}
branch of the quadratic spectrum. This branch corresponds to terms with
the same mode index, \(m=n\), and opposite frequency signs, for which the
spectral center is
\(\omega_m-\omega_n=0\). The corresponding linewidths add, producing a
zero-frequency Lorentzian \(L(2\gamma_m,\omega)\). By contrast, the
sum-frequency branches are centered near \(\pm 2\omega_m\), while the
off-diagonal difference branches are centered near
\(\pm(\omega_m-\omega_n)\), outside the filter passband for the modes
considered here. As shown in Appendix~\ref{app:noise_pathways}, this
gives the low-frequency, high-temperature reduction
\begin{align}
S(\omega)&\approx
S_\eta(\omega)+d^2\sum_m A_m^2 L(2\gamma_m,\omega),\label{eq:reduced_noise_spectrum} 
\end{align}
where $A_m\equiv b_m^2/(\beta M\omega_m^2)$. As a result, the decay exponent can be expressed as
\begin{align}
\chi(\Omega_0,\lambda,d)
&= \frac{1}{2\pi} \int_{0}^\infty d\omega\, F(\omega, \Omega_0,\lambda, T) S_\eta(\omega) \notag\\
&\quad + d^2 \sum_m A_m^2\, I(\gamma_m, \Omega_0,\lambda, T),
\label{eq:chi_decomp}
\end{align}
with a slight abuse of notation to emphasize that the control duration and modulation frequency parameters affect both integrals. The overlap integral between the Lorentzian axial contribution and the control FF is given by
\begin{equation}
I(\gamma, \Omega_0, \lambda, T) \equiv \frac{1}{2\pi}\int_{0}^{\infty} d\omega\,F(\omega, \Omega_0,\lambda, T) L(2\gamma,\omega). \label{eq:I}
\end{equation}
For the DR waveforms used in this study, this overlap expression admits the closed form
\begin{align}
I(\gamma, \Omega_0, \lambda, T)=& \frac{\Omega_0^2}{2(\lambda^2+\gamma^2)^2}\notag\\
&\times
\left[
\gamma T(\lambda^2+\gamma^2)
+2\lambda^2\left(1-e^{-\gamma T}\right)
\right]. 
\label{eq:I-DRFF}
\end{align}

\subsection{Survival Probabilities}

To estimate the control-noise spectrum in the presence of both strong dephasing and control noise, the QNS protocol \cite{Maloney2023, norris2018optimally} uses the measured combination of survival probabilities
\begin{align*}
\mathcal{P}(T)=\frac{1}{2}\bigl[1+P_{X}(T)-P_{Y}(T)-P_{Z}(T)\bigr]. 
\end{align*}
Here, \(P_O(T)\) denotes the probability in which the ion is prepared in the \(+1\) eigenstate of \(O\) and still resides there when measured at time \(T\). 
Using the reduced density matrix in Eq.~\eqref{eq:rhoT}, we find $P_Z(T)=P_Y(T) = (1+e^{-\chi})/2$ and $P_X(T)=1$; thus, 
\begin{equation}
    \mathcal{P}(T)=(1-e^{-\chi})/2.
    \label{eq:P_chi}
\end{equation}
Within the present model, which includes only control noise, there is no obvious advantage to using $\mathcal{P}(T)$ instead of $P_Z(T)$. The benefit becomes apparent when dephasing noise along the dynamically suppressed $Z$ axis is present.
This linear combination cancels the fourth-order dephasing contributions present in the individual survival probabilities and thus isolates the control-noise contribution up to fourth order \cite{Maloney2023}.

\subsection{Effective Decay Parameterization}

From the preceding analysis, Eq.~\eqref{eq:chi_decomp}
separates the decay exponent into a curvature-independent control-noise contribution and a curvature-dependent axial-motion term. For the DR waveforms, the FF scales as $F(\omega,T)\propto \Omega_0^2$, so the decay exponent inherits an overall $\Omega_0^2$ scaling. We therefore write 
\begin{equation}
\chi(\Omega_0,\lambda,d) \equiv \xi(\lambda,d)\,\Omega_0^2 \label{eq:chi_scaling}
\end{equation}
where $\xi$ is independent of $\Omega_0$ and can be decomposed as
\begin{align}
\xi(\lambda,d) &= \xi_{\eta}(\lambda) + d^2\,\xi_{\mathrm{axial}}(\lambda), \\
\xi_{\eta}(\lambda) &= \frac{1}{2\pi} \int_{0}^\infty d\omega\, F(\omega, 1,\lambda, T) S_\eta(\omega) \label{eq:native_overlap} \\
\xi_{\mathrm{axial}}(\lambda) &= \sum_m A_m^2\, I(\gamma_m, 1,\lambda, T)
\end{align}
with $\xi_{\eta}$ capturing the native control-noise contribution and $\xi_{\mathrm{axial}}$ encoding the axial-motion contribution through the local beam curvature.

This form makes clear that, for fixed axial equilibrium position and pulse shape, measurements at different Rabi rates probe the same underlying parameter $\xi$, differing only by the known scaling with $\Omega_0^2$. This allows data taken at multiple drive amplitudes to be combined in a single estimate of $\xi(\lambda, d)$ for each beam position and associated curvature.

\subsection{QNS Spectral Reconstruction}\label{sect:QNS}

The native control-noise spectrum $S_\eta(\omega)$ is not known \emph{a priori}.
However, Eq.~\eqref{eq:native_overlap} shows that each measured
$\xi_\eta(\lambda)$ is a linear functional of $S_\eta(\omega)$. 
Because the DR FFs $F(\omega,1,\lambda,T)$ are peaked at $\omega=\lambda$ but have finite width and sidelobes, each measurement probes a weighted average of $S_{\eta}(\omega)$ over a finite frequency window. This spectral leakage must be accounted for to avoid biasing the reconstruction.

We expand $S_{\eta}(\omega)$ in a discrete basis $\{\phi_k(\omega)\}$ (e.g. piecewise‑constant or linear):
\begin{equation}
S_{\eta}(\omega)=\sum_{k}s_k\,\phi_k(\omega).
\end{equation}
Inserting this into Eq.~\eqref{eq:native_overlap} yields a linear system
\begin{equation}
\xi_{\eta}(\lambda_i)=\sum_{k}W_{ik}s_k,
\end{equation}
where
\begin{equation}
W_{ik}= \frac{1}{2\pi}\int_{0}^{\infty}\!d\omega\,
F(\omega,1,\lambda_i,T)\,\phi_k(\omega).
\end{equation}
The matrix $W$ captures both the intended passband and the leakage.  
The inverse problem’s conditioning is set by the filter bandwidth $\sim2\pi/T$, so we choose piecewise-constant basis functions whose support matches this resolution scale.

\section{Experimental Findings} \label{sec:experiment}
We now employ the protocol described in Sec.~\ref{sec:protocol} to characterize control noise in the presence of axial motion. Experiments are performed on the Quantum Scientific Computing Open User Testbed (QSCOUT) \cite{Clark2021_qscout}, a trapped ion quantum
processor located at Sandia National Laboratories. As shown below, the predictions of our model are in good agreement with the experimental results, enabling estimation of the power spectral density of the laser control noise on the QSCOUT testbed.

\subsection{Platform Description} \label{sect:platform}

The QSCOUT testbed consists of a linear chain of Yb-171 ions in a single harmonic potential is confined via radio frequency (RF) and direct current (DC) fields in a microfabricated surface ion trap. A 370\,nm laser is used to perform Doppler cooling, state preparation and detection of the hyperfine qubit states,  $^2S_{1/2}
|F = 0,m_F = 0\rangle$ and $^2S_{1/2}|F = 1,m_F = 0\rangle$, split by 12.6 GHz.
A pulsed 355\,nm Raman laser perpendicular to the axis of the ion chain drives the qubit transition. By designing the pulses that vary the amplitude in the ways outlined above, we were able to perform the spectroscopic studies of control noise. 
The 355\,nm laser is split into 32 individual-addressing laser beams in a multichannel AOM that allows independent control of the laser frequency, phase and amplitude delivered to each ion in the chain. Raman-driven qubit rotations are implemented in a copropagating geometry by applying two tones to any individual addressing beam. The individual addressing beams are spaced $4.5~\mu m$ apart with roughly a $0.8~\mu m$  $1/e^2$ intensity waist.  
Alignment to the spacing and beam center of the individual addressing laser beams is set by the DC fields that define the trap axial frequency and ion position.  For consistency, experiments were always performed with the axial frequency of $\omega_{\rm axial}/2\pi$ = 0.327\,MHz, which defines a spacing consistent with a 6~ion chain, even when only using a single ion.

\begin{figure}[t]
    \centering
    \includegraphics[width=\linewidth]{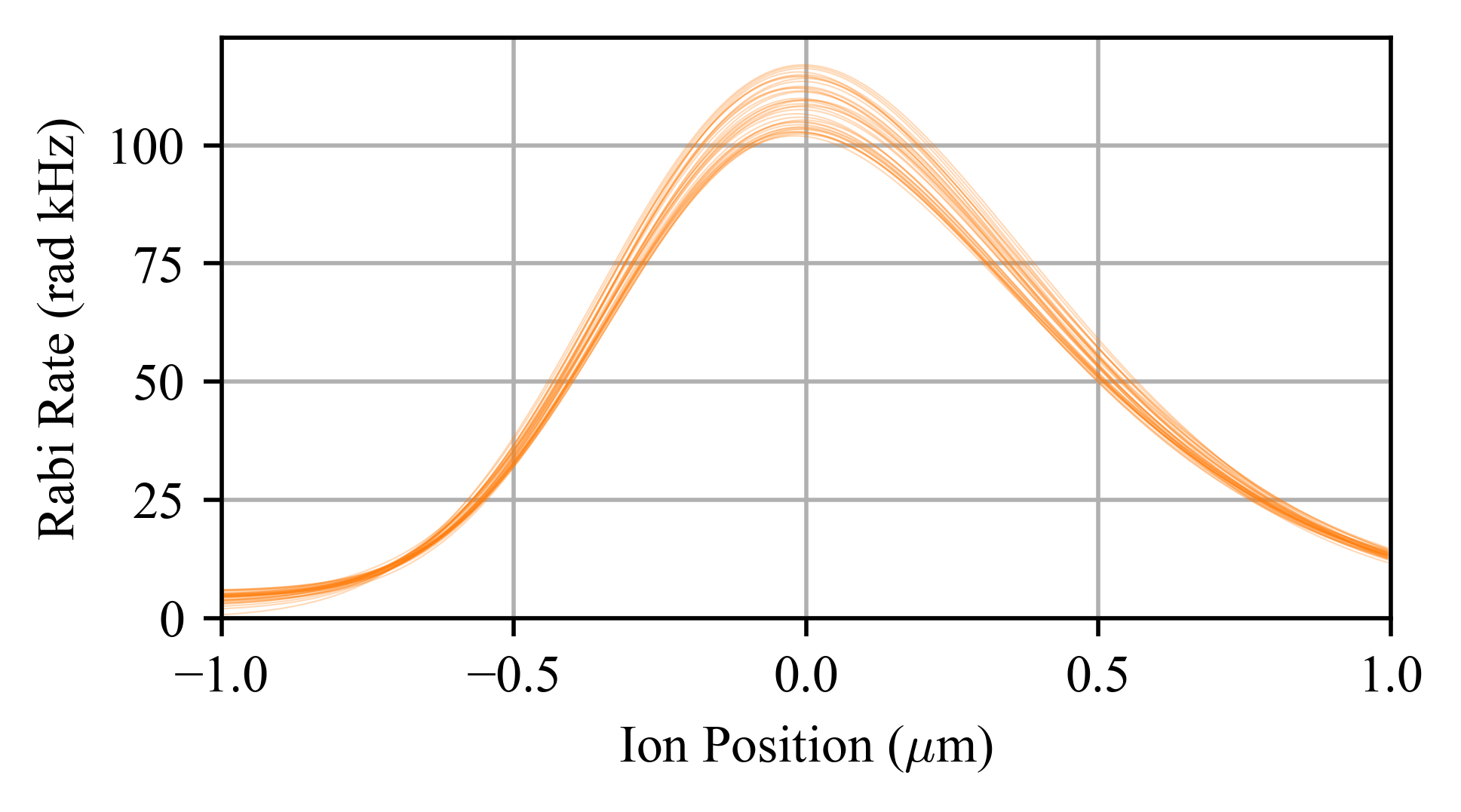}
    \caption{ Beam-profile calibrations used to determine the ion position within the 355\,nm beam. A single ion was displaced along the trap axis, allowing its axial motion to sample different local beam intensities. Skew-Gaussian fits to the resulting scans reveal calibration-to-calibration drift in beam center, peak intensity, and envelope shape, including significant profile asymmetry. }
    \label{fig:beam_profile}
\end{figure}

The QSCOUT platform is particularly suited to the characterization study performed here since it offers access to pulse level control via JaqalPaw \cite{lobserJaqalpaw}. 
For the first set of experiments, a single ion was moved along the trap axis such that it sampled different beam intensities as shown in Fig.~\ref{fig:beam_profile}. Measurements of Rabi oscillations were recorded for different positions of the ion in the beam profile.

In the second set of experiments, we used a fixed pulse duration of 3 ms, and used 9 evenly spaced DR modulation frequencies  $\lambda$ ranging up to 3 kHz. Each $\lambda$ has a discrete family of DR Rabi rates $\Omega_0$ imposed by the condition in Eq.~\eqref{eq::J0Condition} \cite{Maloney2023}; we ran the QNS circuit for all of the available drive amplitudes up to 126 rad kHz. For each ion position, this yields 52 combinations of $(\lambda, \Omega_0)$. With each configuration, we measured the QNS survival probability in the $X$, $Y$, and $Z$ bases. Since the system is always initialized and measured in the Z-basis, for operation in the $X$ and $Y$ bases, we perform $\pi/2$ rotations before and after the DR sequence.  We repeated this measurement for 37 ion positions evenly spaced between $\pm 0.45\;\mu$m away from the beam center, each separated by 25 nm. In total, we measured the survival probability for 5772 different configurations. Due to the large number of configurations, we used 200 shots per experiment.

\subsection{Decay Factors} \label{sect:decay-factors}

The survival probability $\mathcal{P}(T)$ is related to the decay exponent $\chi$ via Eq.~\eqref{eq:P_chi}, and for the DR waveforms used here, the model reduces to $\chi(\Omega_0,\lambda,d) \equiv \xi(\lambda,d)\,\Omega_0^2$ as in Eq.~\eqref{eq:chi_scaling}. Thus, for an experimental class defined by a fixed axial offset and pulse-envelope shape given by DR modulation frequency $\lambda$, the data are fully characterized by a single parameter $\xi(\lambda,d)$. Experimentally, we realize a family of measurements within each class by varying only the overall drive amplitude, and hence the Rabi rate, while holding the beam position and modulation frequency fixed. These measurements probe the same underlying noise environment and beam geometry, differing only by the known $\Omega_0^2$ scaling of the decay.

\begin{figure}[t]
    \centering
    \includegraphics[width=\linewidth]{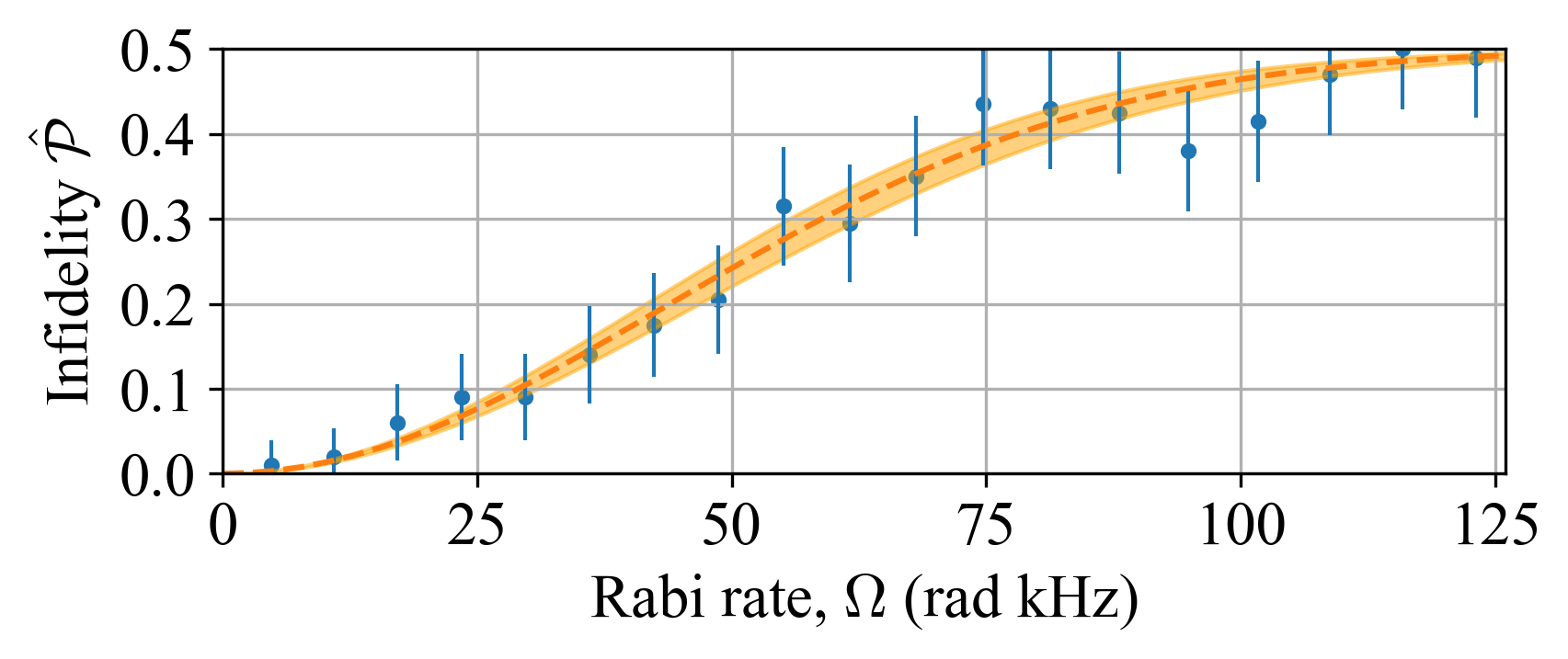}
    \caption{The infidelity curve for $\lambda=0.33$ kHz obtained at an offset of $0.425\;\mu$m, fit according to Eq.~\eqref{eq:curve_fit} with a value of $\xi = 2.6\pm0.3\times10^{-10}\;(\textrm{rad Hz})^{-2}$. Shot-noise intervals are shown in blue calculated from the 95\% Wilson score confidence interval (CI) at each Rabi rate, and the propagated infidelity with the 95\% CI for $\xi$ in orange.
    } \label{fig:infidelity_vs_rabi}
\end{figure}

We estimate $\xi(\lambda, d)$ at each axial offset by jointly maximum-likelihood fitting all measurements at a fixed offset and DR modulation frequency to the binomial model with
\begin{equation}
\hat{\mathcal{P}}(\xi; \Omega_0)=\frac{1}{2}\left(1-e^{-\xi \Omega_0^2}\right), \label{eq:curve_fit}
\end{equation}
where $\hat{\mathcal{P}}$ is the measured probability from the QNS experiments. As the error budget is dominated by shot noise, we neglect other uncertainties (such as measurement error, contributing $< 1\%$ error) in the fit. This global fit enforces the shared $\Omega_0^2$ dependence and aggregates statistics across all Rabi rates, improving the precision of the estimate relative to fits performed using a single $\Omega_0$. A representative example is shown in Fig. \ref{fig:infidelity_vs_rabi}. For some configurations, the infidelity saturates quickly at low Rabi rates; in others the fidelity remains quite high even at the largest drive amplitude sampled.

\begin{figure*}[t]
    \centering
    \includegraphics[width=1\linewidth]{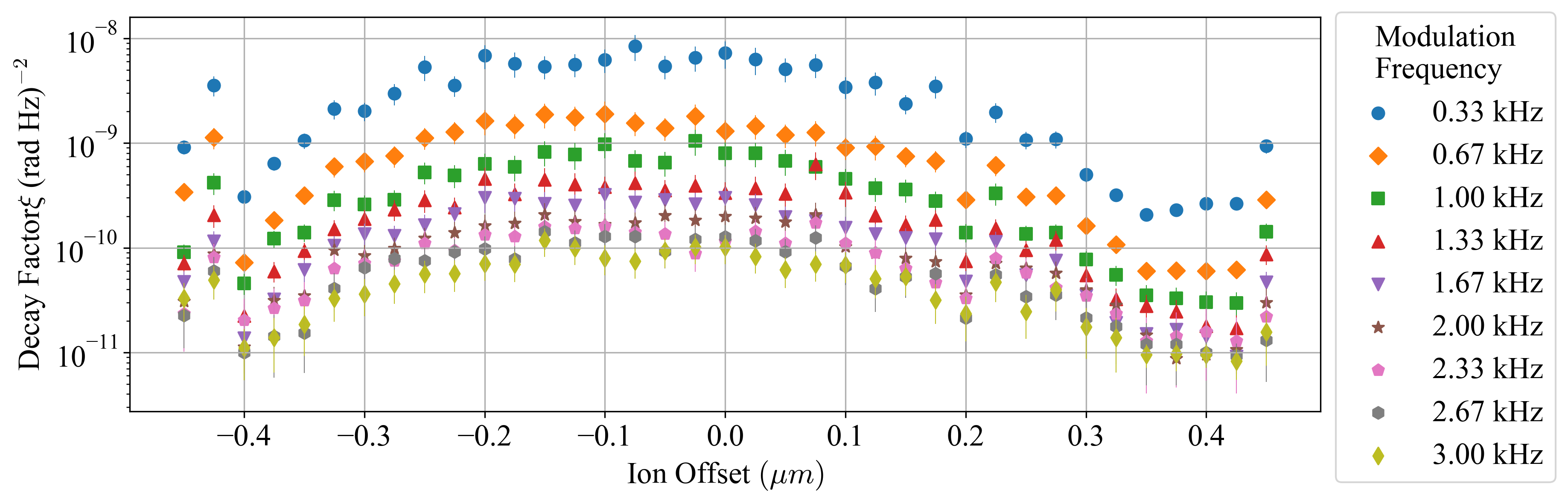}
    \caption{
Decay factors extracted from DR-QNS measurements as a function of axial ion offset from the beam center.
Each point shows the fitted value of \(\xi\) obtained from the shared-amplitude model
\(\chi=\xi\Omega_0^2\) for a fixed axial offset and DR modulation frequency \(\lambda\).
Error bars indicate the fit uncertainty from binomial shot noise.
The decay factors vary strongly with both offset and modulation frequency: low-frequency DR waveforms show the largest decay, while higher-frequency waveforms are increasingly suppressed.
The broad position dependence reflects conversion of axial motion into effective amplitude noise through the local beam curvature, with increased decay near the beam center where the quadratic coupling is large. }
    \label{fig:xi-label}
\end{figure*}

The resulting decay factors are shown in Fig.~\ref{fig:xi-label}. The extracted values span approximately one order of magnitude as a function of axial offset and two orders of magnitude across the sampled modulation frequencies, indicating a strong dependence of the effective noise on both position and frequency. Within the model, the position variation is attributed to the curvature-dependent axial-motion contribution $d^2\,\xi_{\mathrm{axial}}(\lambda)$: near points of small beam curvature, the conversion of axial motion into amplitude noise is suppressed, while away from these points the same motion leads to significantly enhanced decay.
The residual offset at minimal curvature reflects the curvature-independent control-noise contribution $\xi_\eta$.

The frequency dependence of the axial component is governed by the overlap integrals $I(\gamma_m, \Omega_0, \lambda, T)$. Inspecting Eq.~\eqref{eq:I}, this overlap is generally largest at low $\lambda$; the salient feature is how rapidly it drops off with modulation frequency. In the model, this falloff is controlled by the linewidths $\gamma_m$, with a sharper drop indicating narrower axial spectra. The observed trend where $\xi$ is almost $4\times$ larger for 0.33 kHz than for 0.67 kHz is consistent with axial motion characterized by narrow linewidths $\gamma_m <$ 0.33 kHz, as supported by the detailed model fitting shown in the next section.

\subsection{Beam Profile Characterization} \label{sect:data-analysis}

The axial-motion contribution to the measured decay depends on the local curvature of the Raman beam intensity profile sampled by the ion. Estimating this curvature was therefore essential for comparing the measured decay factors to the axial-motion model. In practice, however, accurately characterizing the beam profile proved nontrivial due to both spatial asymmetry in the beam shape and slow temporal drift over the duration of the experiment.

Prior calibrations on the QSCOUT system had shown that the individual addressing beams were not well described by symmetric Gaussian profiles. Instead, the central region of the beam was consistently asymmetric and was more accurately captured by a phenomenological skewed-profile model. In addition, we observed fine structure on the left side of the beam, occasionally appearing as small secondary peaks in the measured Rabi-rate scans. Over the several-hour duration of the experiment, the beam position also drifted relative to the ion(s). Consequently, a nominal axial offset specified through the trap control fields could not be assumed to correspond to a fixed physical location within the beam profile throughout data collection.

To track and mitigate these effects, we performed a beam-profile calibration prior to each QNS configuration. Each calibration consisted of a position sweep of the ion across the beam profile in which the ion was displaced to 61 axial positions spanning a total range of \(2~\mu\mathrm{m}\). At each position, we applied a laser pulse with fixed duration and amplitude and estimated the excitation probability from 250 experimental shots. The scan range was chosen to extend well beyond the \(\pm0.45~\mu\mathrm{m}\) window used for the QNS measurements themselves in order to sample the full support of the beam profile, including regions where the achievable Rabi rate was too small to implement the DR drive amplitudes. Because the beam exhibited asymmetry, fine structure, and slow positional drift over the duration of the experiment, calibrations restricted to the narrower operating region produced unstable estimates of the beam center and local curvature. Sampling the broader $2$ $\mu$m support of the beam provided sufficient global information to consistently track beam drift and extract local curvature estimates within the smaller QNS measurement window.

Rather than fitting all scans to a single static beam profile, we treated each position sweep independently using an empirical skew-Gaussian parameterization, as  shown in Fig.~\ref{fig:beam_profile}. To compensate for slow beam drift without overfitting individual scans, the beam center and overall amplitude scale were allowed to vary between calibrations, while the width and skewness parameters were constrained near their initially calibrated values. The fitted beam center from each scan was then used to determine the physical location of the target QNS offset for that experimental configuration. As a result, although the targeted offsets used in the experiment were nominally spaced by \(25~\mathrm{nm}\), the inferred absolute axial positions after drift correction were not exactly uniformly spaced. Additionally, sporadic features near $-0.4~\mu\mathrm{m}$ were visible in the raw scans, indicating residual beam-profile structure not captured by the skew-Gaussian ansatz.

\begin{figure}[b]
    \centering
    \includegraphics[width=\linewidth]{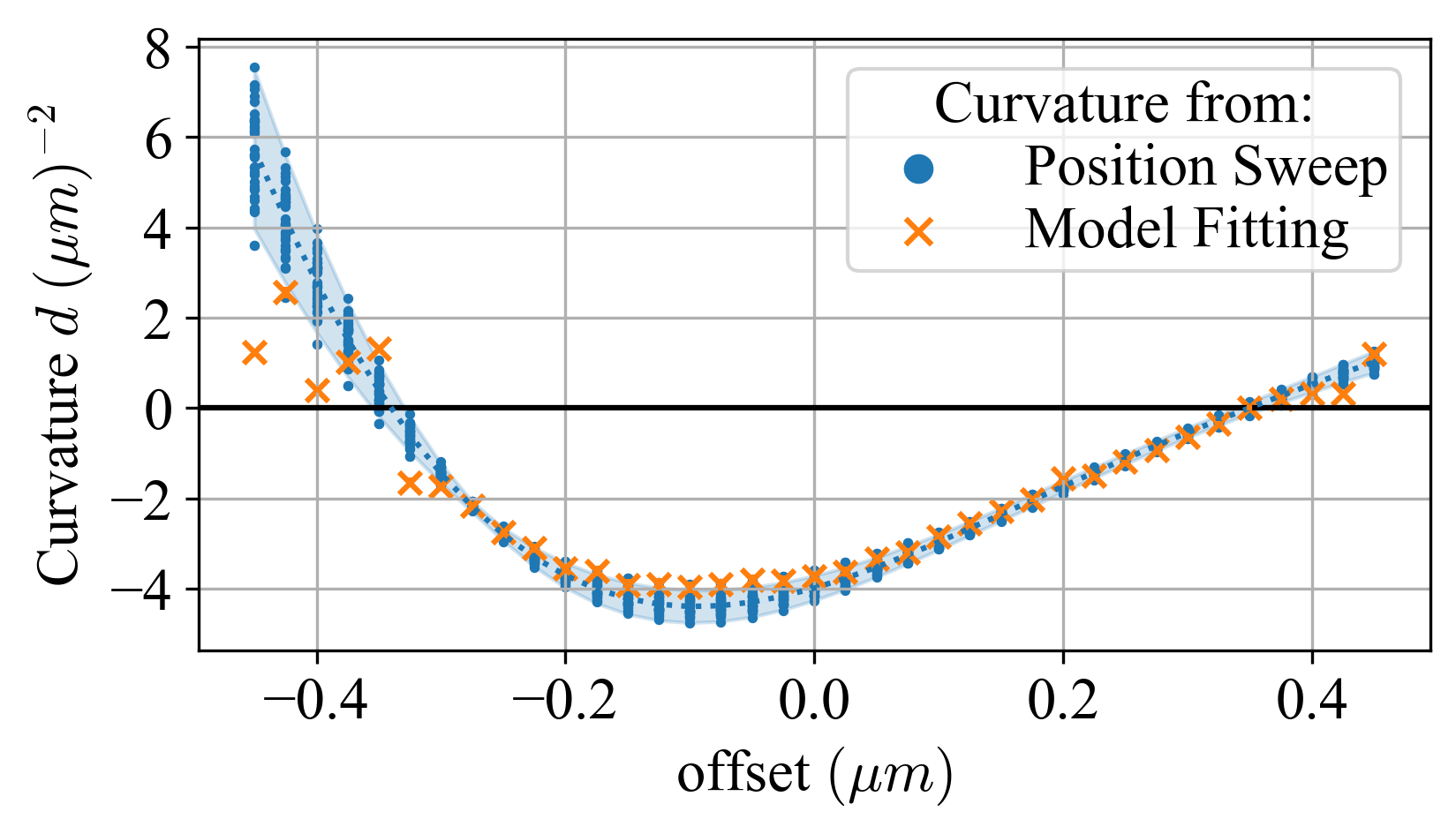}
\caption{Beam-profile curvature used in the axial-motion noise model. Blue points show curvatures extracted from repeated position-sweep calibrations at drift-corrected ion positions; the dotted line and shaded band show the mean and empirical 95\% CI. Orange crosses show curvatures inferred from the global model fit. Near-zero curvature indicates beam inflection points, where quadratic coupling of axial motion to amplitude noise is suppressed.}
\label{fig:curvature}
\end{figure}

\begin{figure*}[t]
    \centering    \includegraphics[width=\linewidth]{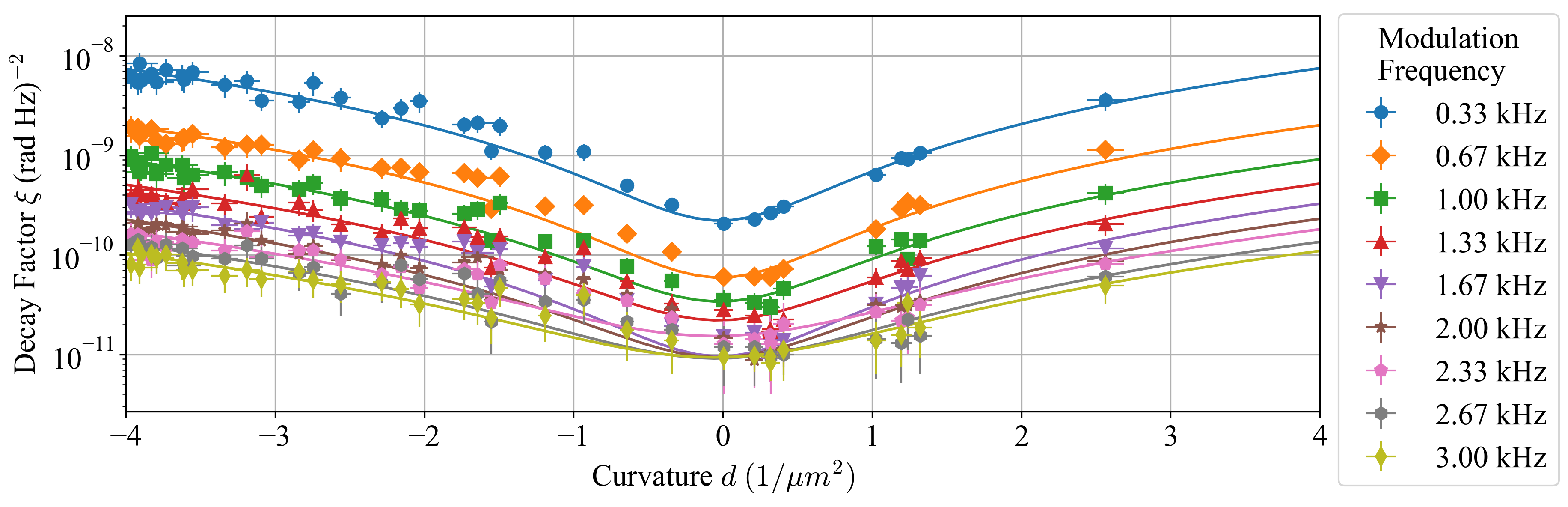}
    \caption{ Global fit of the axial-motion noise model to experimentally extracted DR-QNS decay factors. Points show $\xi(\lambda,d)$ versus local beam curvature $d$ for each DR modulation frequency $\lambda$, and solid curves show the fitted model. The fit includes a curvature-independent native control-noise contribution and a curvature-dependent axial-motion contribution proportional to $d^2$. The model captures the measured decay factors across nearly three orders of magnitude in $\xi$. The minimum near $d=0$ across DR waveforms is consistent with suppression of quadratic conversion of axial motion into effective amplitude noise at the beam inflection point. }
\label{fig:model_fit}
\end{figure*}

From each independently fitted beam profile, we computed the local second derivative of the intensity profile and used this quantity as the effective curvature parameter entering the axial-motion model. Because the extracted curvatures depended sensitively on the fitted beam shape, particularly near regions exhibiting fine structure, we did not treat any individual fit as exact. Instead, we regarded the collection of independently calibrated beam profiles as an ensemble estimate of the underlying curvature. At each axial position, we therefore calculated the mean and standard deviation of the inferred curvatures across all calibrations and used these as the empirical curvature estimate and associated uncertainty in the subsequent model fitting. This procedure partially mitigates the effects of shot noise, finite scan resolution, and slow temporal drift, although it does not fully eliminate systematic modeling error arising from deviations of the true beam profile.

To account for these uncertainties during parameter estimation, we employed Deming regression \cite{deming1943statistical} when fitting the axial-motion model to the experimentally extracted decay factors. Rather than treating the inferred curvatures as exact independent variables, the regression simultaneously optimized corrected curvature values together with the global model parameters by minimizing the normalized deviations in both the measured decay factors and the curvature estimates. This distinction proved important for several data points near the far left side of the beam profile, where the global fit consistently favored smaller positive curvatures than those predicted by the skew-Gaussian ensemble alone. This behavior is consistent with the aberrant fine structure observed experimentally in that region of the beam and suggests that the phenomenological beam model systematically overestimated the local curvature there. Nevertheless, these data points still contributed useful information to the global fit because the relative dependence of the measured decay factors on the DR waveform parameters continued to constrain the underlying axial-motion and control-noise parameters even when the local curvature estimate itself was imperfect.

\subsection{Center of Mass Mode Characterization} \label{sect:CoM-characterization}

Combining the estimates of the curvature and decay factors, we fit the remaining free parameters in the model. We obtain good agreement with experiment across all orders of magnitude. We obtain a center of mass (COM) occupation number of $\langle n_0\rangle = 205 \pm 49$ corresponding to a temperature of $0.32 \pm 0.08$ mK, and for the linewidth, we get $\gamma = 0.165 \pm 0.112 \,\text{kHz}$. Both thermal axial motion and native laser noise were significant. We also consider control noise not from axial motion, which presents as the decay factor at zero curvature. This is analyzed more in the next subsection.

Independently, the temperature of the axial mode was measured by adding an additional laser beam to the setup that addresses a quadrupole transition (as opposed to the hyperfine transition used for the rest of this work) with a propagation direction that can access the motion in the axial direction.  We estimate the occupation of the axial COM mode $\langle n_0\rangle $ to be $120$~quanta by extrapolating from measurements at higher confinement to the inter-ion spacing used in this work, corresponding to an axial frequency of $0.327\times 2\pi$~MHz.  We do not place an error bar on the occupation estimated by this method, since we believe that would give a misleading amount of precision as the extrapolation was determined from a limited number of points and the frequency dependence is not certain from literature. However, our estimated value is order-of-magnitude consistent with the value calculated above.  The difficulty in obtaining an estimate for the axial occupation underscores the need for new methods to extract this value, as offered in this work. Additionally, an independent measure of the axial thermal linewidth is not possible in our system due to the large linewidth of the laser driving the quadrupole transition.  

These results clarify the tradeoff between gate speed and motion-induced control noise. Operating near the beam center maximizes Rabi rate but increases sensitivity to axial motion through beam curvature, while operating near an inflection point suppresses this contribution at reduced drive strength. On QSCOUT, this tradeoff corresponds to roughly a factor-of-two reduction in Rabi rate for an order-of-magnitude reduction in control noise.

\subsection{Reconstructed Control-Noise Spectrum} \label{sect:control-spectra}

\begin{figure}[t]
    \centering    \includegraphics[width=\linewidth]{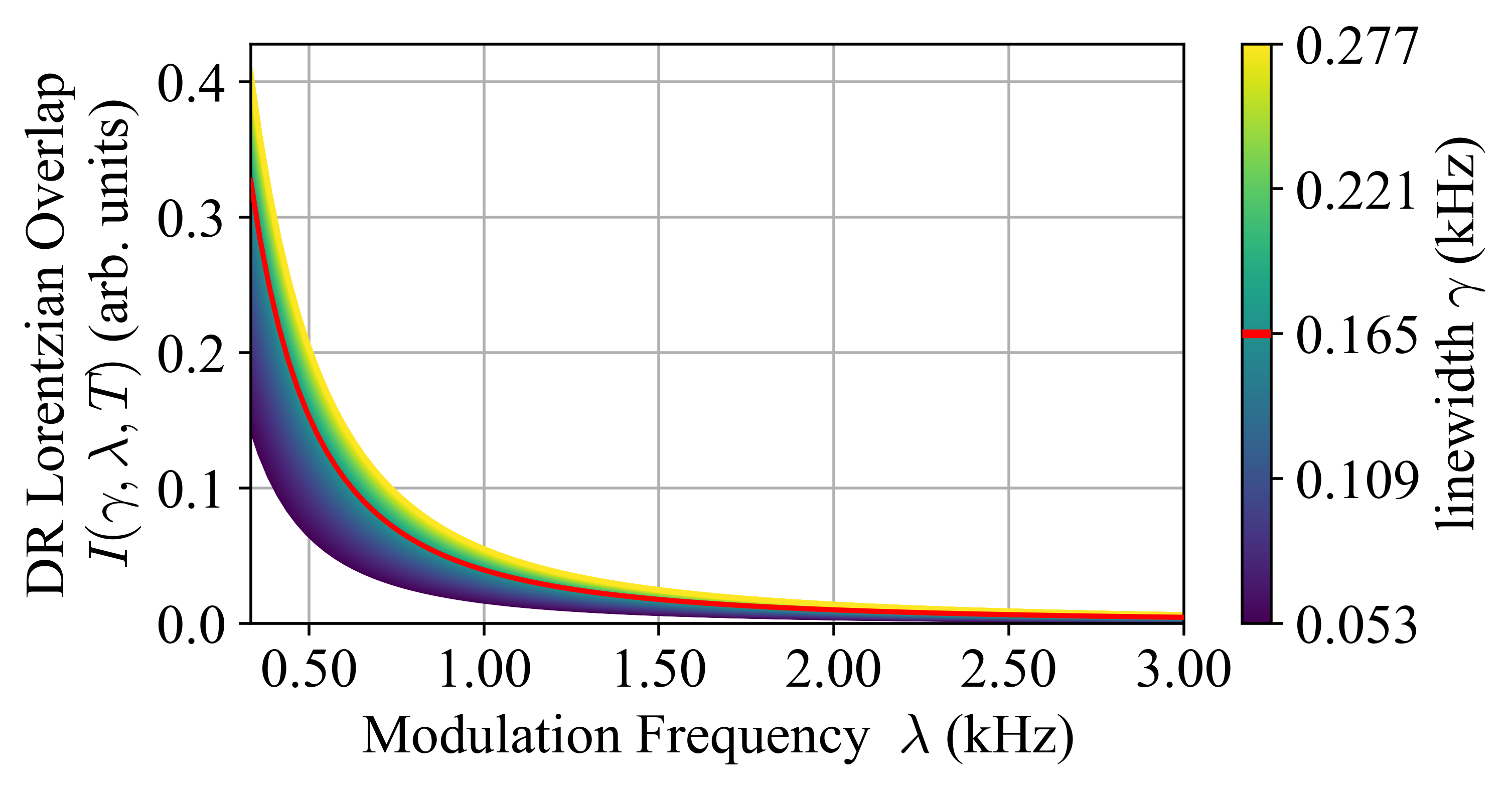}
    \caption{ DR filter overlap with the thermal Lorentzian axial-motion spectrum, Eq.~\eqref{eq:I}. The overlap $I(\gamma,\lambda,T)$ sets the axial-motion contribution to the DR-QNS decay for linewidth $\gamma$ and interpolated modulation frequency $\lambda$. Colors indicate the fitted linewidth uncertainty range, while the red curve shows the best-fit value $\gamma = 0.165~\mathrm{kHz}$. The overlap falls off rapidly with $\lambda$, indicating that the axial-motion contribution is concentrated at low modulation frequencies. }
    \label{fig:dr_lorentzian}
\end{figure}

Using the QNS inversion framework described in Section~\ref{sect:QNS}, we reconstruct the native control-noise spectrum $S_\eta(\omega)$ from the measured overlap coefficients $\xi_\eta(\lambda)$. We parameterize the spectrum as piecewise constant over frequency intervals defined by the midpoints between adjacent DR modulation frequencies used in the experiment. This choice yields spectral bins naturally centered on the nine sampled modulation frequencies while maintaining a frequency resolution comparable to the DR filter bandwidth. The lowest- and highest-frequency bins are extended to $0$ and $\infty$, respectively. The resulting spectral weights are estimated using non-negative least-squares inversion.

\begin{figure}[t]
    \centering
    \includegraphics[width=\linewidth]{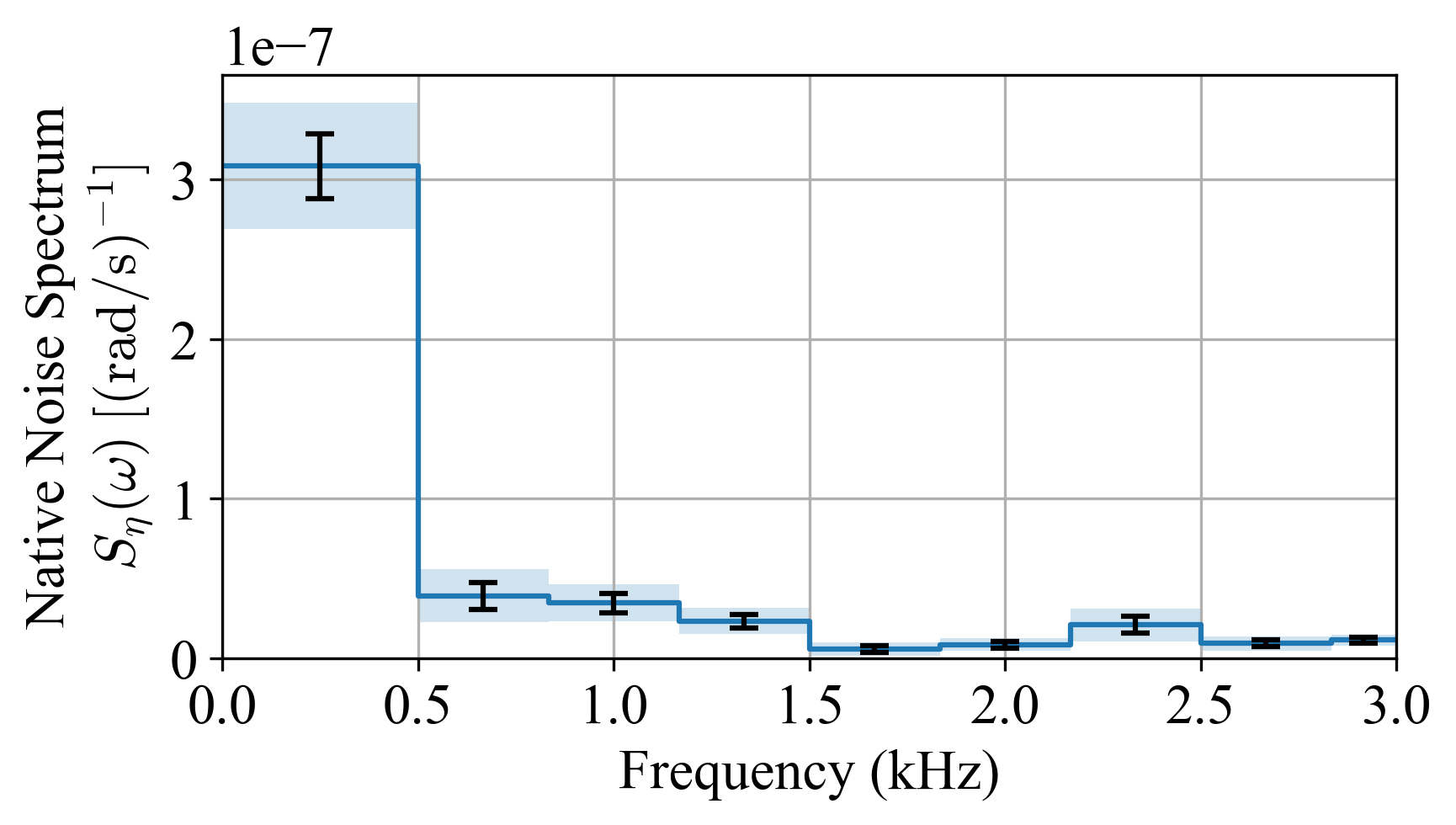}
    \caption{Reconstructed native control-noise spectrum obtained from the curvature-independent DR-QNS decay component. The spectrum is represented by piecewise-constant frequency bins centered on the sampled modulation frequencies and estimated by non-negative least-squares inversion. The blue band shows the 95\% CI, and black error bars indicate the propagated standard deviation. The spectral weight is concentrated at low frequencies, consistent with slowly varying native amplitude noise. }
    \label{fig:spectra}
\end{figure}

Figure~\ref{fig:spectra} shows the reconstructed spectrum. The spectral weight is concentrated at low frequencies, indicating that the residual control noise varies slowly on the timescale of the control sequence. In this regime, standard mitigation techniques such as dynamical decoupling or composite pulses are expected to be effective, as they are designed to suppress quasi-static low-frequency amplitude fluctuations.

We also examine the curvature-dependent axial contribution, which is an order-of-magnitude stronger than the native noise at the beam center. In Fig.~\ref{fig:dr_lorentzian} we show the FF overlap from Eq.~\eqref{eq:I} for the range of linewidths obtained from model fitting, and interpolated between DR modulation frequencies. As expected for a narrow linewidth, the overlap rapidly drops off with higher frequency, particularly for $\lambda > 1$ kHz. This suggests that other control methods for suppressing control noise, such as pulse shaping \cite{2022TroutOptimalControl} or sufficiently rapid dynamical decoupling, may be effective on this platform if they inhibit the response below this frequency.

\subsection{Multi-ion chain} \label{sect:multi-ion}

\begin{figure*}[ht]
    \centering\includegraphics[width=0.85\linewidth]{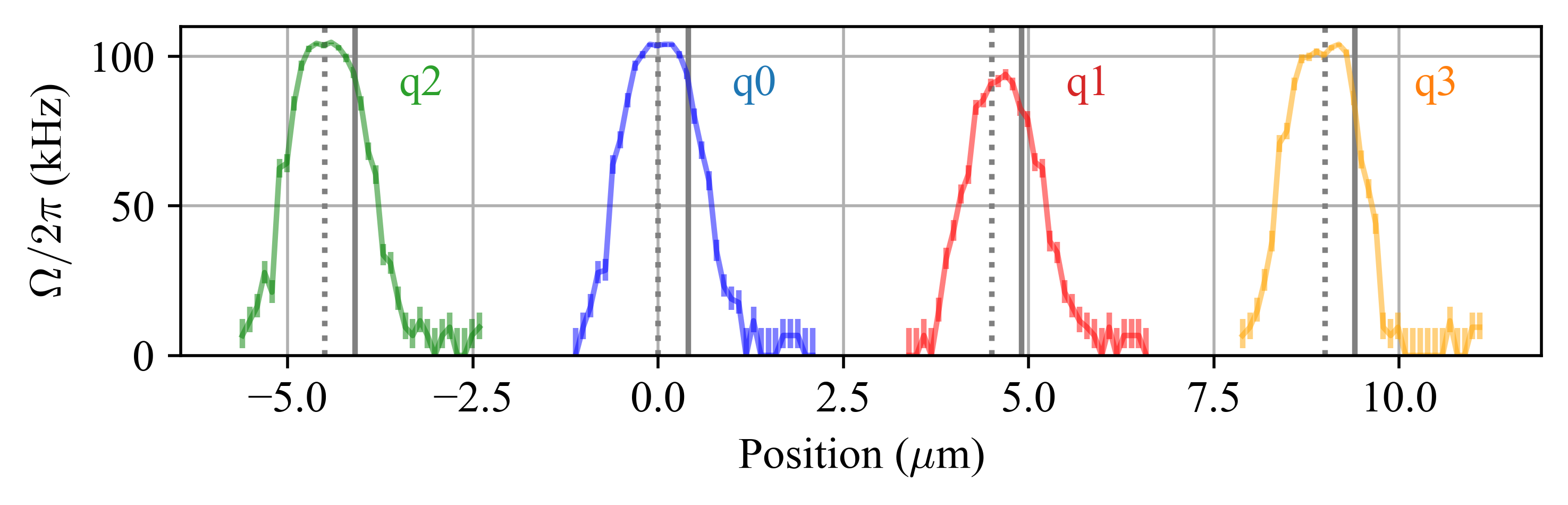}
    \includegraphics[width=0.85\linewidth]{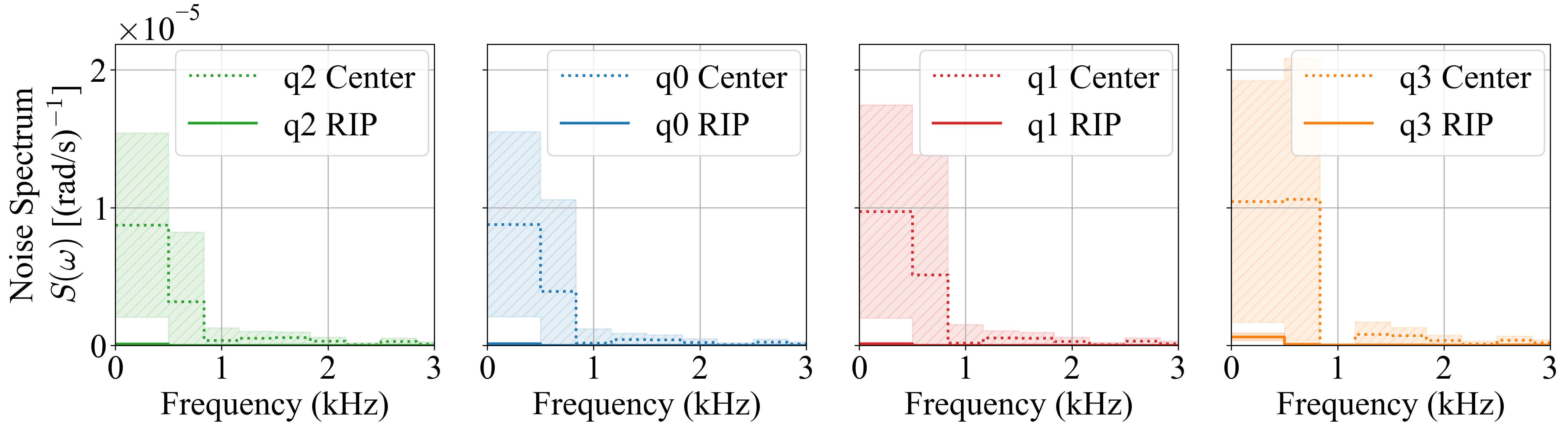}
   
    \caption{Alignment calibration data (top) and noise spectra (bottom) of four data ions in four individual addressing beams. 
    (top) A six-ion chain with a central, four data ion subregister is moved along the trap axis and Rabi rate of the individual addressing beams is measured.  The relative ion positions used to align with each beam center are marked with a vertical dashed line, and a vertical solid line marks the right inflection point positions.
    (bottom) QNS is performed at both the beam center location (dashed) and with the chain position offset to align to a point of near zero curvature (the right-hand side inflection point, RIP) of the beam. The contribution of axial motion to control errors as measured by QNS is substantially suppressed at the inflection point for all qubits.
    }
    \label{fig:four_ions}
\end{figure*}

To confirm that our findings from the single-ion studies apply to multi-qubit registers, we perform the same spectroscopy protocol in parallel on a four-data-ion register within a six-ion chain. In this multi-ion configuration, we characterize the control noise spectrum both at the beam center, and at the inflection point of the beam, where curvature is ideally zero and the impacts of axial motion are mitigated. We find that the reconstructed noise spectra at the inflection point demonstrates significantly reduced control errors as compared to the beam center (as shown in Fig.~\ref{fig:four_ions}), consistent with our single-ion findings. The same beam-position tradeoff observed for a single ion therefore extends to an entire register.

These results complement those of Ref.~\cite{Monroe2022}, which characterized the effect of axial motion on Rabi oscillations under constant drive. Whereas Ref.~\cite{Monroe2022} inferred motional effects through the phase advance and contrast decay of Rabi oscillations, here we characterize the underlying temporal correlations of this process by reconstructing the effective control-noise spectrum using QNS. This extends axial-motion characterization from average coherence loss to frequency-resolved control-noise spectroscopy.

\section{Conclusion}
We have shown that axial motion in an individually addressed trapped-ion processor can be characterized through the effective amplitude noise induced as ions sample a transverse laser-beam profile. Building on the model for axial-motion-induced control noise introduced in Ref.~\cite{Monroe2022}, we use DR-QNS to move beyond constant-drive measurements and probe the frequency dependence of this noise process. This allows curvature-independent native amplitude control noise to be separated from curvature-dependent axial-motion noise, while employing the same transverse Raman beams used for qubit control rather than a dedicated axial spectroscopy beam.

The measured curvature and frequency dependence are well described by a model in which thermal axial motion is converted into low-frequency amplitude noise through the local beam curvature. The global fit captures the measured DR-QNS decay factors across nearly three orders of magnitude and gives motional parameters comparable to an independent sideband-based temperature estimate. This demonstrates that DR-QNS can provide access not only to the presence of axial-motion-induced control noise, but also to the spectral structure of the underlying axial dynamics, including parameters that are difficult to measure directly on the present platform.

We further demonstrated that the DR-QNS protocol can be applied in parallel across four ions in a chain. In this setting, operating near the beam inflection point suppresses the curvature-mediated noise contribution for several individually addressed beams simultaneously. As a result, the ion-position tradeoff observed in single-ion measurements persists at the register level. The parallel DR-QNS measurements provide a scalable diagnostic for identifying beam positions where motion-induced control noise is reduced across an ion chain.

More broadly, curvature-resolved DR-QNS provides a route to characterizing native control-noise mechanisms that depend on ion position and motional dynamics. Extending these measurements to larger registers could enable spatial mapping of correlated control noise while providing a valuable diagnostic for optimizing beam alignment and mitigating motion-induced errors in trapped-ion processors. By enabling parallel characterization of axial-motion-induced control noise, DR-QNS potentially provides a scalable approach for investigating how spatially correlated control errors propagate through multi-qubit operations and ultimately impact quantum algorithms and fault-tolerant protocols.

\begin{acknowledgments}

This work was in part supported by the U.S. Department of Energy, Office of Science, Office of Advanced Scientific Computing Research, Accelerated Research in Quantum Computing under Award Number DE-SC0020316 and DE-SC0025509.  This research was also supported [in part] by the U.S. Department of Energy, Office of Science, Office of Advanced Scientific Computing Research Quantum Testbed Program and by Sandia National Laboratories' Laboratory Directed Research and Development Program. Sandia National Laboratories is a multi-mission laboratory managed and operated by National Technology \& Engineering Solutions of Sandia, LLC (NTESS), a wholly owned subsidiary of Honeywell International Inc., for the U.S. Department of Energy’s National Nuclear Security Administration (DOE/NNSA) under contract DE-NA0003525. SAND2026-22647O.  This written work is authored [in part] by an employee of NTESS. The employee, not NTESS, owns the right, title and interest in and to the written work and is responsible for its contents. Any subjective views or opinions that might be expressed in the written work do not necessarily represent the views of the U.S. Government. The publisher acknowledges that the U.S. Government retains a non-exclusive, paid-up, irrevocable, world-wide license to publish or reproduce the published form of this written work or allow others to do so, for U.S. Government purposes. The DOE will provide public access to results of federally sponsored research in accordance with the DOE Public Access Plan.

This manuscript was prepared with the assistance of AI-based writing tools for typesetting and language refinement. The authors have reviewed and are responsible for the final content.
\end{acknowledgments}

\begin{appendix} \label{appendix}

\section{Derivation of Axial Mode Correlation Function} \label{app:Lindblad}

We model the axial gain function by Taylor expanding the beam profile to second order about the ion's equilibrium position, $f(x)\propto a+b x+d x^2$.  For a beam-intensity profile \(I(x)\) expanded about the ion's equilibrium position \(x=0\), the coefficients are proportional to the local value and derivatives of the profile \[
a=I(0),\qquad b=I'(0) , \qquad d=\frac{1}{2}I''(0), \] up to the overall normalization of \(f(x)\). Thus, \(b\) captures the first-order sensitivity to displacement, and \(d\) describes the local curvature of the beam profile.

We choose the coefficients so that the mean axial gain satisfies $\langle f\rangle=1$. Assuming $\langle x\rangle=0$, the normalized gain can be written in terms of the thermal displacement variance $\sigma^2\equiv\langle x^2\rangle$ as
\begin{equation}
    f(x)
    =
    \frac{a}{a+d\sigma^2}
    +
    \frac{b}{a+d\sigma^2}x
    +
    \frac{d}{a+d\sigma^2}x^2,
\end{equation}
where the rescaled coefficients depend on the thermal variance and are therefore temperature dependent. This normalization sets the Rabi rate by the effective mean-field laser intensity and ensures that the beam envelope \(\Omega(t)\) retains the expected calibrated scale despite axial fluctuations, which corrects for what Ref.~\cite{Monroe2022} called \emph{phase advance}. In our experiments, the phase advance adjusted the curvatures by around 4\%. Throughout the rest of this paper, we reuse \(b\) and \(d\) to denote these normalized linear and quadratic coefficients.

We begin with a quantum description of the axial motion of an ion in a chain. The displacement operator $\hat{x}$ for a particular ion can be expressed as in Eq.~\eqref{eq:displacement}.
As described in Sec.~\ref{subsec:qubit-axial-coupling}, the axial modes are described by the bare harmonic oscillator additionally coupled to a thermal environment. Under the standard assumptions of weak-coupling and negligible back-action, the evolution of the axial modes under the influence of the external environment can be described by a Lindblad master equation $\dot{\rho}=\mathcal{L}[\rho]$, where
\begin{eqnarray}
\mathcal{L}[\cdot]
= \sum_m \Big(
&-i&[\omega_m(\hat{a}_m^\dagger \hat{a}_m + \tfrac{1}{2}), \cdot] \nonumber\\
&+& \gamma_m(n_m+1)
(\hat{a}_m \cdot \hat{a}_m^\dagger
- \tfrac{1}{2}\{\hat{a}_m^\dagger \hat{a}_m, \cdot\}) \nonumber\\
&+& \gamma_m n_m
(\hat{a}_m^\dagger \cdot \hat{a}_m
- \tfrac{1}{2}\{\hat{a}_m \hat{a}_m^\dagger, \cdot\})
\Big).
\label{eq:lindblad}
\end{eqnarray}
$\gamma_m$ is the damping rate or linewidth of the axial mode-environment coupling, $\beta$ is the inverse temperature, and $n_m\equiv 1 \big/ (e^{\beta \hbar \omega_m} - 1)$ is the thermal quanta, which in the high temperature limit is approximately $1\big/\beta \hbar \omega_m$. As is well-known~\cite{breuer2002theory}, the steady state for this Lindbladian is the thermal Gibbs state $\rho_{\textrm{th}} \propto \exp(-\beta H_a)$, which is Gaussian in position because $H_a$ is quadratic.

We can solve Eq.~\eqref{eq:lindblad} in the Heisenberg picture for the annihilation operators $\dot{\hat{a}}_m = \mathcal{L}^\dag [\hat{a}_m]$ and obtain the time-dependent displacement operator $x(t)$ for $t\ge0$ with $\hat{a}_m(t)= \hat{a}_m e^{- i \omega_m t - \gamma_m t / 2}$.

This leads to a correlation function 
\begin{eqnarray}
C_x(t) &=& \sum_m \frac{\hbar b_m^2}{2M\omega_m}
e^{-\gamma_m \lvert t \rvert /2} \times\nonumber\\
&&\,\Bigl[(2n_m+1)\cos(\omega_m t)-i\sin(\omega_m t)\Bigr].
\end{eqnarray}
For \(t\ge0\), the two-time correlations are obtained from the quantum regression theorem, which states that they evolve under the same Lindblad propagator as the corresponding single-time expectation values. The result is extended to \(t<0\) using stationarity and \(C_x(-t)=C_x(t)^*\), yielding the factor \(e^{-\gamma_m|t|/2}\).

\section{Calculating the Cumulant} \label{app:Keldysh_cumulant}

We derive Eq.~\eqref{eq:logKpm} starting from the off-diagonal coherence factor $K_{+-}(T)$ [see Eq.~\eqref{eq:decay-factor}]
\begin{equation}
K_{+-}(T)
=
\left\langle
\mathcal T_-
e^{-i\int_0^Tdt\,\frac{\Omega(t)}{2}\tilde f(t)}
\mathcal T_+
e^{-i\int_0^Tdt\,\frac{\Omega(t)}{2}\tilde f(t)}
\right\rangle.
\end{equation}

We expand the two exponentials directly to second order and retain the
second cumulant. Since
\(\langle \tilde f(t)\rangle=0\), the first-order terms vanish, giving
\begin{align}
\log K_{+-}(T)
=
-\frac{1}{8}
\int_0^T dt_1
\int_0^T dt_2\,
\Omega(t_1)\Omega(t_2)\,
\mathcal{A}(t_1,t_2),
\label{eq:four_cases_corrected}
\end{align}
where
\begin{align}
\mathcal{A}(t_1,t_2)
={}&
\left\langle
\mathcal{T}_+
\tilde f(t_1)\tilde f(t_2)
\right\rangle
+
\left\langle
\mathcal{T}_-
\tilde f(t_1)\tilde f(t_2)
\right\rangle
\nonumber\\
&+
\left\langle
\tilde f(t_1)\tilde f(t_2)
\right\rangle
+
\left\langle
\tilde f(t_2)\tilde f(t_1)
\right\rangle.
\label{eq:correlator}
\end{align}
The first two terms arise from contractions within the forward and
backward branches, respectively, while the final two terms arise from
contractions between the two branches.

We now simplify the bracketed expression. By definition of time-ordering and anti-time-ordering, the first term in $\mathcal{A}(t_1,t_2)$ can be expressed as 
\begin{align}
\langle \mathcal T_+ \tilde f(t_1)\tilde f(t_2)\rangle
=&\;
\Theta(t_1-t_2)\langle \tilde f(t_1)\tilde f(t_2)\rangle \notag\\
&+\Theta(t_2-t_1)\langle \tilde f(t_2)\tilde f(t_1)\rangle,
\end{align}
while the second term is
\begin{align}
\langle \mathcal T_- \tilde f(t_1)\tilde f(t_2)\rangle
=&\;
\Theta(t_2-t_1)\langle \tilde f(t_1)\tilde f(t_2)\rangle \notag\\
&+\Theta(t_1-t_2)\langle \tilde f(t_2)\tilde f(t_1)\rangle.
\end{align}
Adding these two relations and using
$
\Theta(t_1-t_2)+\Theta(t_2-t_1)=1
$
gives
\begin{multline}
\langle \mathcal T_+ \tilde f(t_1)\tilde f(t_2)\rangle
+
\langle \mathcal T_- \tilde f(t_1)\tilde f(t_2)\rangle
= \\\left\langle
\tilde f(t_1)\tilde f(t_2)
\right\rangle
+
\left\langle
\tilde f(t_2)\tilde f(t_1)
\right\rangle.
\label{eq:tplus_tminus_identity}
\end{multline}
Substituting Eq.~\eqref{eq:tplus_tminus_identity} into Eq.~\eqref{eq:correlator}, we obtain
\begin{equation}
\mathcal{A}(t_1,t_2)
=
2\left\langle
\left\{\tilde f(t_1),\tilde f(t_2)\right\}
\right\rangle.
\end{equation}
Defining the symmetrized covariance as in Eq.~\eqref{eq:C_sym},
\[
\tilde C(t_1-t_2)
\equiv
\frac{1}{2}
\left\langle
\left\{
\tilde f(t_1),
\tilde f(t_2)
\right\}
\right\rangle,
\]
we have
\[
\mathcal A(t_1,t_2)=4\tilde C(t_1-t_2).
\]
Substitution into Eq.~\eqref{eq:four_cases_corrected} yields
Eq.~\eqref{eq:logKpm}. The decomposition of \(\tilde C(t)\) into
laser-amplitude and axial-motion correlations is derived in
Appendix~\ref{app:effective_gain_correlation}.

\section{Spectrum Derivation}
\label{app:spectra_derivation}

\subsection{Effective-Gain Correlation Function}
\label{app:effective_gain_correlation}

We first express the cumulant function in
Eq.~\eqref{eq:C_sym} in terms of the laser-amplitude noise and axial
displacement cumulants. The normalized axial gain may be written as
\begin{equation}
f(\hat x(t))
=
1+b\hat x(t)
+d\left[\hat x^2(t)-\sigma^2\right],
\label{eq:normalized_gain_app}
\end{equation}
where \(b\) and \(d\) denote the normalized linear and quadratic
coefficients as described in Appendix~\ref{app:Lindblad}. Due to stationarity, $\langle\hat x^2\rangle \equiv \langle \hat x^2(t) \rangle = \langle \hat x^2(0) \rangle$ ensuring $f(\hat x(t))$ is always normalized. We therefore define the zero-mean axial-gain fluctuation
\begin{equation}
\delta f_x(t)
\equiv
f(\hat x(t))-1
=
b\hat x(t)
+d\left[\hat x^2(t)-\sigma^2\right].
\label{eq:delta_fx_app}
\end{equation}
Using $f(\hat x(t))=1+\delta f_x(t)$, the zero-mean total effective-gain fluctuation defined in
Eq.~\eqref{eq:H-tilde-f} is
\begin{align}
\tilde f(t)
&\equiv
[1+\eta(t)]f(\hat x(t))-1 \\
&=
\eta(t)+\delta f_x(t)+\eta(t)\delta f_x(t).
\label{eq:delta_ftilde_decomp_app}
\end{align}

We assume that the laser-amplitude noise and axial motion are independent
stationary processes. The laser fluctuation satisfies
\(\langle\eta(t)\rangle=0\), while
\(\langle\delta f_x(t)\rangle=0\) follows from the definition
\(\delta f_x(t)\equiv f(\hat{x}(t))-1\) and the normalization
\(\langle f(\hat{x}(t))\rangle=1\). Since \(\eta(t)\) is classical and commutes with the axial
operators, substitution of Eq.~\eqref{eq:delta_ftilde_decomp_app} into
Eq.~\eqref{eq:C_sym} gives
\begin{equation}
\tilde C(\tau)
=
C_\eta(\tau)
+
C_f(\tau)
+
C_\eta(\tau)C_f(\tau),
\label{eq:Ctilde_decomp_app}
\end{equation}
where \(\tau\) is the time difference $t_1 - t_2$ and
\begin{align}
C_\eta(\tau)
&\equiv
\left\langle\eta(\tau)\eta(0)\right\rangle,
\label{eq:Ceta_app}\\
C_f(\tau)
&\equiv
\frac{1}{2}
\left\langle
\left\{
\delta f_x(\tau),\delta f_x(0)
\right\}
\right\rangle.
\label{eq:Cf_app}
\end{align}
The mixed terms between \(\eta\) and \(\delta f_x\) vanish because the
two processes are independent and individually have zero mean. The final
term in Eq.~\eqref{eq:Ctilde_decomp_app} arises from the multiplicative
fluctuation \(\eta(t)\delta f_x(t)\).
For zero-mean Gaussian axial motion, all odd displacement moments vanish.
The terms in \(C_f(\tau)\) proportional to \(bd\) therefore vanish, leaving
\begin{equation}
C_f(\tau)
=
b^2 C_x^{\mathrm{sym}}(\tau)
+
d^2 C_{x^2}^{\mathrm{sym}}(\tau),
\label{eq:Cf_axial_decomp_app}
\end{equation}
where
\begin{align}
C_x^{\mathrm{sym}}(\tau)
&\equiv
\frac{1}{2}
\left\langle
\left\{
\hat x(\tau),\hat x(0)
\right\}
\right\rangle,
\label{eq:Cx_sym_def_app}\\
C_{x^2}^{\mathrm{sym}}(\tau)
&\equiv
\frac{1}{2}
\left\langle
\left\{
\hat x^2(\tau)-\sigma^2,
\hat x^2(0)-\sigma^2
\right\}
\right\rangle.
\label{eq:Cx2_sym_def_app}
\end{align}

The symmetrized displacement correlation is the real part of the
unsymmetrized correlation in Eq.~\eqref{eq:Cxt},
\begin{equation}
C_x^{\mathrm{sym}}(\tau)
=
\operatorname{Re}C_x(\tau).
\label{eq:Cx_sym_relation_app}
\end{equation}
Applying Wick's theorem to the Gaussian axial state gives the exact
quadratic covariance
\begin{equation}
C_{x^2}^{\mathrm{sym}}(\tau)
=
2\operatorname{Re}\!\left[C_x(\tau)^2\right].
\label{eq:Cx2_exact_app}
\end{equation}
In the high-temperature regime relevant to the experiment, the contribution
from the displacement commutator is suppressed relative to the thermal
symmetrized fluctuations by \(\mathcal{O}(n_m^{-2})\). We therefore use
\begin{equation}
C_{x^2}^{\mathrm{sym}}(\tau)
\simeq
2\left[C_x^{\mathrm{sym}}(\tau)\right]^2,
\label{eq:Cx2_highT_app}
\end{equation}
so that Eq.~\eqref{eq:Cf_axial_decomp_app} becomes
\begin{equation}
C_f(\tau)
\simeq
b^2 C_x^{\mathrm{sym}}(\tau)
+
2d^2\left[C_x^{\mathrm{sym}}(\tau)\right]^2.
\label{eq:Cf_highT_app}
\end{equation}

Combining Eqs.~\eqref{eq:Ctilde_decomp_app} and
\eqref{eq:Cf_highT_app}, the correlation function entering the cumulant
is therefore
\begin{equation}
\begin{aligned}
\tilde C(\tau)
={}&
C_\eta(\tau)
+b^2 C_x^{\mathrm{sym}}(\tau)
+2d^2\left[C_x^{\mathrm{sym}}(\tau)\right]^2
\\
&+
b^2 C_\eta(\tau)C_x^{\mathrm{sym}}(\tau)
+
2d^2 C_\eta(\tau)
\left[C_x^{\mathrm{sym}}(\tau)\right]^2.
\end{aligned}
\label{eq:Ctilde_explicit_app}
\end{equation}

Using the Fourier-transform convention
\begin{equation}
S_j(\omega)
=
\int_{-\infty}^{\infty}d\tau\,
e^{-i\omega\tau}C_j(\tau),
\label{eq:fourier_convention_app}
\end{equation}
and defining
\begin{align}
S_x(\omega)
&\equiv
\mathcal F\!\left[C_x^{\mathrm{sym}}\right](\omega),
\\
S_{xx}(\omega)
&\equiv
\mathcal F\!\left[
\left(C_x^{\mathrm{sym}}\right)^2
\right](\omega), \label{eq:Sxx_def_app}
\end{align}
the Fourier transform of Eq.~\eqref{eq:Ctilde_explicit_app} appearing in Eq.~\eqref{eq:full_noise_decomp} is
\begin{equation}
\begin{aligned}
S(\omega)
={}&
S_\eta(\omega)
+b^2S_x(\omega)
+2d^2S_{xx}(\omega)
\\
&+
\frac{b^2}{2\pi}
\left(S_\eta*S_x\right)(\omega)
+
\frac{2d^2}{2\pi}
\left(S_\eta*S_{xx}\right)(\omega).
\end{aligned}
\label{eq:full_noise_decomp_app}
\end{equation}
Here we have used the convolution theorem,
\begin{equation}
\mathcal F[C_1C_2](\omega)
=
\frac{1}{2\pi}
(S_1*S_2)(\omega).
\end{equation}
Equation~\eqref{eq:full_noise_decomp_app} is the spectral decomposition
given in Eq.~\eqref{eq:full_noise_decomp} of the main text.

\subsection{Axial Displacement and Quadratic Spectra}
\label{app:axial_spectra}

We now evaluate the axial spectra appearing in
Eq.~\eqref{eq:full_noise_decomp_app}. The symmetrized displacement
correlation is
\begin{align}
C_x^{\rm sym}(t)
&=
\sum_m
\frac{\hbar b_m^2}{2M\omega_m}
(2n_m+1)
e^{-\gamma_m |t|/2}
\cos(\omega_m t),
\label{eq:Cx_sym_app}
\end{align}
as follows from Eqs.~\eqref{eq:Cxt} and
\eqref{eq:Cx_sym_relation_app}. Its Fourier transform defines the
displacement spectrum,
\begin{equation}
S_x(\omega)
=
\int_{-\infty}^{\infty}dt\,
e^{-i\omega t}C_x^{\rm sym}(t).
\end{equation}
Using the Lorentzian convention in
Eq.~\eqref{eq:lorentzian_spectra}, for which
\(L(\gamma,\omega)\) has total area \(2\pi\), and the
high-temperature approximation
\((n_m+\frac{1}{2})\approx 1/(\beta\hbar\omega_m)\), we obtain
\begin{align}
S_x(\omega)
&=
\sum_m
\frac{b_m^2}{2\beta M\omega_m^2}
\left[
L(\gamma_m,\omega-\omega_m)
+
L(\gamma_m,\omega+\omega_m)
\right]
\nonumber\\
&\equiv
\sum_m
\frac{A_m}{2}
\sum_{\pm}
L(\gamma_m,\omega\pm\omega_m),
\label{eq:Sx_app}
\end{align}
with $A_m\equiv b_m^2/(\beta M\omega_m^2)$ as defined in Sec.~\ref{subsec:noise-pathways}.

From Eq.~\eqref{eq:Sxx_def_app}, the quadratic axial spectrum is the
self-convolution of the displacement spectrum,
\begin{equation}
S_{xx}(\omega)
=
\frac{1}{2\pi}(S_x*S_x)(\omega).
\label{eq:Sxx_conv_app}
\end{equation}
Using the Lorentzian convolution identity
\begin{align}
\frac{1}{2\pi}\int_{-\infty}^{\infty}d\nu\,
&L(\gamma_1,\nu-\omega_1)
L(\gamma_2,\omega-\nu-\omega_2)\nonumber\\
=\;
&L(\gamma_1+\gamma_2,\omega-\omega_1-\omega_2),
\label{eq:L_conv_app}
\end{align}
we obtain
\begin{equation}
S_{xx}(\omega)
=
\sum_{m,n}\frac{A_mA_n}{4}
\sum_{\pm,\pm}
L(\gamma_m+\gamma_n,\omega\pm\omega_m\pm\omega_n).
\label{eq:Sxx_app}
\end{equation}

\subsection{Low-Frequency Reduction and Mixed Terms}
\label{app:noise_pathways}

We now identify the terms in the full effective spectrum that contribute
appreciably to the filter overlap in Eq.~\eqref{eq:FFF}. Treating the
laser-amplitude noise and axial motion as independent stationary Gaussian
processes, the effective spectrum is given by Eq.~\eqref{eq:full_noise_decomp}.
For the DR waveforms used here, \(F(\omega,T)\) is concentrated in the
kHz band, whereas the axial mode frequencies satisfy
\(\omega_m/2\pi\sim\mathrm{MHz}\). We also assume that the native
laser-amplitude noise \(S_\eta(\omega)\) is concentrated at low
frequencies. Thus only spectral weight near \(\omega=0\) contributes
appreciably to the filter overlap.

The linear axial contribution,
\begin{equation}
b^2S_x(\omega)
=
b^2\sum_m \frac{A_m}{2}
\sum_{\pm}L(\gamma_m,\omega\pm\omega_m),
\end{equation}
is centered at \(\omega=\pm\omega_m\), well outside the DR passband, and
is therefore negligible. The mixed linear term
\(b^2(S_\eta*S_x)/(2\pi)\) is negligible for the same reason:
convolution with low-frequency \(S_\eta\) broadens the peaks in \(S_x\),
but does not shift them to \(\omega\simeq0\).

For the quadratic spectrum in Eq.~\eqref{eq:Sxx_app}, the sum branches
centered at \(\pm(\omega_m+\omega_n)\) also lie far outside the passband.
The difference branches centered at \(\pm(\omega_m-\omega_n)\) are
likewise off-resonant for \(m\neq n\) for the six-ion chain considered
here \cite{James1998-nt}. The only low-frequency quadratic contribution is
therefore the diagonal difference branch with \(m=n\) and opposite signs,
which gives
\begin{equation}
S_{xx}(\omega)
\;\to\;
\frac{1}{2}\sum_m A_m^2 L(2\gamma_m,\omega).
\label{eq:Sxx_lowfreq_app}
\end{equation}
Retaining this contribution in \(2d^2S_{xx}\), while neglecting the
off-resonant linear and non-diagonal quadratic branches, gives the
reduced spectrum in Eq.~\eqref{eq:reduced_noise_spectrum}.

It remains to justify neglecting the mixed quadratic contribution $2d^2 (S_\eta*S_{xx})(\omega)/2\pi$.
Using the same low-frequency reduction in
Eq.~\eqref{eq:Sxx_lowfreq_app}, the corresponding contribution to the
decay exponent is approximately
\begin{align}
\chi_{\rm mix}
&\equiv
\frac{1}{2\pi}\int_0^\infty d\omega\,
F_\Omega(\omega,T)
\frac{2d^2}{2\pi}
(S_\eta*S_{xx})(\omega)
\nonumber\\
&\approx
d^2\sum_m A_m^2
\frac{1}{(2\pi)^2}
\int_0^\infty d\omega\,
F_\Omega(\omega,T)
(S_\eta*L_m)(\omega),
\label{eq:chi_mix_start_app}
\end{align}
where \(L_m(\omega)\equiv L(2\gamma_m,\omega)\). Using commutativity of convolution and the evenness
of \(F_\Omega\), \(S_\eta\), and \(L_m\),
\begin{equation}
\chi_{\rm mix}
\approx
d^2\sum_m A_m^2
\frac{1}{(2\pi)^2}
\int_0^\infty d\omega\,
(F_\Omega*L_m)(\omega)
S_\eta(\omega).
\label{eq:chi_mix_rewrite_app}
\end{equation}
Thus the mixed term has the same structure as the native control-noise
contribution,
\begin{equation}
\chi_\eta
=
\frac{1}{2\pi}\int_0^\infty d\omega\,
F_\Omega(\omega,T)S_\eta(\omega),
\end{equation}
except that the filter is replaced by the broadened filter
\((F_\Omega*L_m)/(2\pi)\).

For the DR waveforms used here, \(F_\Omega(\omega,T)\) is concentrated in
a narrow passband around \(\lambda\), while \(L_m(\omega)\) is centered at
zero with width \(2\gamma_m\). Since \(L_m\) has total area \(2\pi\), the
normalized kernel \(L_m/(2\pi)\) acts as a local averaging operation.
When \(\gamma_m\) is small compared with the frequency scale over which
\(F_\Omega\) or \(S_\eta\) varies across the passband,
\begin{equation}
\frac{1}{2\pi}(F_\Omega*L_m)(\omega)
\approx
F_\Omega(\omega,T),
\end{equation}
or equivalently,
\begin{equation}
\frac{1}{2\pi}(S_\eta*L_m)(\omega)
\approx
S_\eta(\omega).
\end{equation}
Under this approximation,
\begin{equation}
\chi_{\rm mix}
\approx
d^2\sum_m A_m^2\,\chi_\eta.
\label{eq:chi_mix_factorized_app}
\end{equation}
The mixed term therefore produces a small multiplicative renormalization
of the native control-noise contribution rather than a distinct spectral
signature. In the experimental regime considered here, this correction is
below \(1\%\) of \(\chi_\eta\). By contrast, the retained non-mixed axial
term,
\begin{equation}
\chi_{\rm axial}
=
d^2\sum_m A_m^2
\frac{1}{2\pi}\int_0^\infty d\omega\,
F_\Omega(\omega,T)L(2\gamma_m,\omega),
\end{equation}
contains the linewidth-dependent structure used to model the
curvature-dependent axial-motion contribution. We therefore absorb the
mixed quadratic term into a small renormalization of the native
control-noise contribution and retain the explicit axial term in
Eq.~\eqref{eq:reduced_noise_spectrum}.

\end{appendix}

\bibliographystyle{apsrev4-1}
\bibliography{bibliography}

\end{document}